\documentclass[aps,prb,twocolumn,longbibliography]{revtex4-2}
\usepackage{hyperref}
\hypersetup{
    colorlinks = true,
    linkcolor = [rgb]{0.70, 0.13, 0.13},
    citecolor = [rgb]{0.25, 0.41, 0.88},
    urlcolor = [rgb]{0.25, 0.41, 0.88}
}
\usepackage{amsmath}
\usepackage{amsfonts}
\usepackage{amssymb}
\usepackage{pifont}
\usepackage{mathtools}
\usepackage{bm}
\usepackage{cases}
\usepackage{braket}
\usepackage[version=4]{mhchem}
\usepackage{graphicx}
\allowdisplaybreaks%
\usepackage{makecell}

\usepackage{blindtext}

\makeatletter
\newcommand*{\sumcirclearrowleft}{%
  \DOTSB
  \mathop{
    \mathchoice
      {\rlap{\kern.25em\rotatebox[origin=c]{-90}{$\circlearrowleft$}}{\sum}}
      {\vcenter{\rlap{\kern.2em\rotatebox[origin=c]{-90}{$\scriptscriptstyle\circlearrowleft$}}}{\sum}}
      {\sum}{\sum}
  }\slimits@
}

\newcommand*{\sumcirclearrowright}{%
  \DOTSB
  \mathop{
    \mathchoice
      {\rlap{\kern.25em\rotatebox[origin=c]{90}{$\circlearrowright$}}{\sum}}
      {\vcenter{\rlap{\kern.2em\rotatebox[origin=c]{90}{$\scriptscriptstyle\circlearrowright$}}}{\sum}}
      {\sum}{\sum}
  }\slimits@
}
\makeatother

 \graphicspath{ {./images/} }
 \usepackage{physics}
 
\begin{document}
\title{Quantum Spin Supersolid as a precursory Dirac Spin Liquid\\ in a Triangular Lattice Antiferromagnet}
\author{Haichen Jia$^{2}$}
\thanks{These authors contributed equally to this work.}
 
\author{Bowen Ma$^{2}$}
\thanks{These authors contributed equally to this work.}
  
\author{Zidan Wang$^{2,3}$}
\email{zwang@hku.hk}
 
\author{Gang Chen$^{1,4,5}$}
\email{gangchen.physics@gmail.com}

\affiliation{$^{1}$International Center for Quantum Materials, 
School of Physics, Peking University, Beijing 100871, China}
\affiliation{$^{2}$Department of Physics and HK Institute of Quantum Science \& Technology,  
The University of Hong Kong, Pokfulam Road, Hong Kong, China}
\affiliation{$^{3}$Quantum Science Center of Guangdong-Hong Kong-Macau Great Bay Area, 3 Binlang Road, Shenzhen, China}
\affiliation{$^{4}$The University of Hong Kong Shenzhen Institute of Research and Innovation, 
Shenzhen 518057, China}
\affiliation{$^{5}$Collaborative Innovation Center of Quantum Matter, 100871, Beijing, China}

\date{\today}

\begin{abstract}
Based on the recent experiments on the triangular lattice antiferromagnet Na$_2$BaCo(PO$_4$)$_2$,
we propose the easy-axis XXZ spin-1/2 model on the triangular lattice, 
that exhibits a quantum spin supersolid, to be a precursory Dirac spin liquid. 
Despite the presence of a three-sublattice magnetic order as a spin supersolid, 
we suggest that this system is close to a Dirac spin liquid by exploring its spectroscopic response. 
The physical consequence is examined from the spectroscopic response, and 
we establish the continuous spectra near the M point in addition to the K point excitation 
from the spinon continuum on top of the three-sublattice order. 
Moreover, the satellite peaks were predicted at the mid-points connecting the $\Gamma$ and K points. 
This proposal offers a plausible understanding of the recent inelastic neutron scattering measurement 
in Na$_2$BaCo(PO$_4$)$_2$ and could inspire further research in relevant models and materials,
such as K$_2$Co(SeO$_3$)$_2$ and Rb$_2$Co(SeO$_3$)$_2$, and even more 
anisotropic magnets like PrMgAl$_{11}$O$_{19}$. 
\end{abstract}

\maketitle

\section{Introduction} 
\label{sec1}

The U(1) Dirac spin liquid (DSL) has been viewed as the mother state of 
the competing orders such as the N\'{e}el and valence bond solid states in 2D~\cite{PhysRevB.72.104404}.   
By studying the symmetry properties of the monopole operators of the emergent quantum electrodynamics  
and ensuring the concomitant of the ordering-driven spontaneous mass generation and the confinement, 
recent theoretical work suggested the increased stability of the U(1) DSL on frustrated lattices~\cite{DSLMotherState}. 
It inspires that the signatures of DSL may occur even 
in the nearby ordered phases on frustrated lattices.  
In this work, we attempt to identify the specific material and model that may 
reveal the signatures of DSL and competing orders on the triangular lattice.

Recent experiments on the triangular lattice antiferromagnet Na$_2$BaCo(PO$_4$)$_2$ 
show several pieces of puzzling phenomena~\cite{sheng2024continuum}. While the system is clearly ordered 
antiferromagnetically with a three-sublattice order of the spin supersolid type, 
the magnetic excitations do not behave like typical spin-wave modes. Instead, broad  
continuous magnetic excitations appear in both the momentum and energy domains. 
Moreover, there exist strong continuous excitations at low energies around the M 
points that seem to be quite unexpected for an antiferromagnetic order at the K point.   
We interpret these continuous excitations as the signatures for the precursory U(1) DSL.

\begin{figure}[b]
\centering
\includegraphics[width=0.48\textwidth]{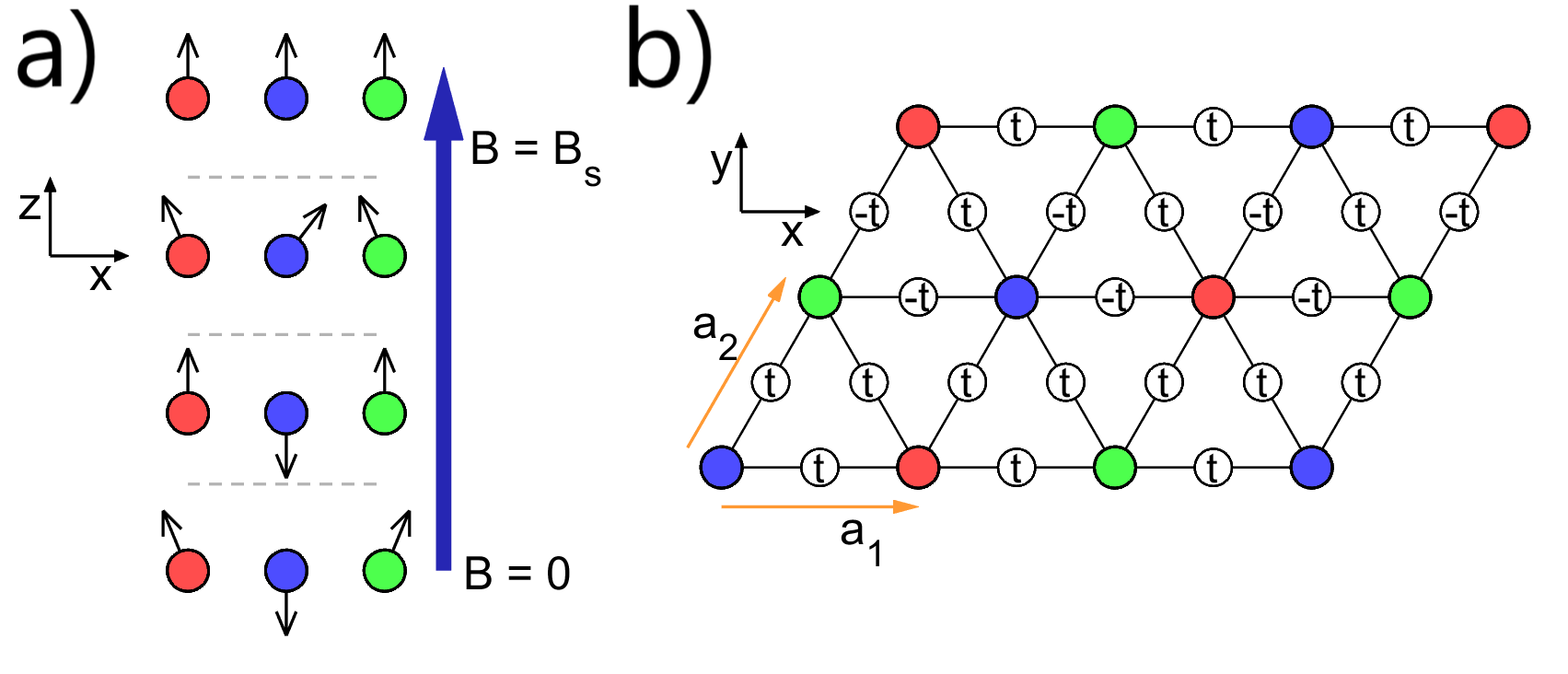}
\caption{ 
(a) The spin orientation of various magnetic phases for the easy-axis XXZ model on the triangular lattice. 
As the field ${{\mathbf{B}} = B \hat{z}}$ increases from zero, the system enters the 3-sublattice 
supersolid (Y-shape), UUD, V-shape and the fully polarized states. 
The sublattices are plotted in red, green and blue.
(b) The spinon hoppings $t_{ij}$ are chosen with a gauge choice supporting a $\pi$ flux. } 
\label{fig: fig1}
\end{figure}

For the more material's aspect of the physics, it is the conventional belief that    
the $3d$ transition metal oxides should have more symmetric spin models.  
In reality, all the $3d$ transition metal ions have rather different physical properties, 
exhibiting rich behaviors with a diverse of models that depend sensitively on the structures 
and the crystal field environments~\cite{PhysRevLett.102.096406,PhysRevB.96.020412,PhysRevB.100.045103,Su_2019}. 
In many cases, the orbitals get involved and wildly change the physics. For the Co ion, 
the often-discussed scenario is actually about the selection of the high or low spin states 
due to the competition between the Hund's coupling and the crystal electric field splitting. 
More recently, however, the spin-orbit coupling of the Co ion has received quite some attention. 
This occurs when the electrons partially fill the $t_{2g}$ shell~\cite{Witczak_Krempa_2014}. 
Due to the active spin-orbit coupling and the orbitals, the local moment of the relevant Co ion is 
no longer pure spins, and is instead an entangled object of spin and orbital. 
The resulting effective model is expected to be anisotropic instead of Heisenberg-like. 
The well-known limiting examples were the quasi-1D Ising magnets CoNb$_2$O$_6$, 
BaCo$_2$V$_2$O$_8$ and 
SrCo$_2$V$_2$O$_8$~\cite{CoNb2O6, BaCo2V2O8, SrCo2V2O8,PhysRevLett.127.077201,PhysRevLett.120.207205,Wang_2018}. 
For the same reason, the Kitaev interaction has been extended to the honeycomb cobaltates including
Na$_2$Co$_2$TeO$_6$ and Na$_2$Co$_2$SbO$_6$~\cite{PhysRevLett.125.047201,PhysRevB.97.014407,PhysRevB.97.014408},
and a similar form of Kitaev interaction is re-invoked for CoNb$_2$O$_6$~\cite{Morris_2021,PhysRevB.105.224421}. 
For our motivation of the triangular lattice antiferromagnet Na$_2$BaCo(PO$_4$)$_2$
and other Co-based triangular lattice compounds, the model is expected
to be the spin-1/2 easy-axis XXZ model~\cite{sheng2022two,sheng2024continuum}.

Na$_2$BaCo(PO$_4$)$_2$ develops a three-sublattice antiferromagnetic order below about 0.15K. 
The weak exchange couplings allow the external magnetic fields about 2T 
to polarize the spins for both in-plane and out-of-plane directions~\cite{NBCP_magnetization}. 
The magnetic excitations of these polarized states are {\sl quantitatively} captured 
by the linear spin-wave theory based on the spin-1/2 XXZ model with strong fields. 
The ratio of the exchange coupling for the $xy$ components over the $z$-component coupling 
is ${\sim 0.58}$, and both are antiferromagnetic and thus {\sl fully-frustrated}. 
The spin-1/2 XXZ model on the triangular lattice is a well-studied problem. 
The phase diagrams with and without the field are quite clear.  
The supersolid order, that breaks both the U(1) symmetry via the in-plane spin 
order and the lattice translation via the $z$-component order, 
was known to be present in the ferromagnetic $xy$ and the antiferromagnetic $z$ regime, 
and was later shown to persist to the fully-frustrated regime with both antiferromagnetic 
$xy$ and $z$ couplings until the Heisenberg point~\cite{XXZ_phase_diagram,PhysRevLett.102.017203}
and is entirely due to quantum origin. Therefore, the antiferromagnetic order
in Na$_2$BaCo(PO$_4$)$_2$ at zero field is a quantum spin supersolid at low temperatures.  
Due to the U(1) symmetry and the easy-axis coupling, the intermediate field along the $z$   
direction stabilizes a 1/3 magnetization plateau with a `UUD' (``up-up-down'') spin configuration
in Na$_2$BaCo(PO$_4$)$_2$. This is consistent with the expectation of the XXZ model. 
What is not understood from the model is the presence of large continuous magnetic excitations 
at zero field. This feature is absent for Ba$_3$CoSb$_2$O$_9$ that is described by  
the same model but in the easy-plane regime~\cite{Kazuhisa,PhysRevLett.116.087201}. 
This difference is expected to arise from the enhanced frustration in the easy-axis regime 
of the XXZ model.

To understand the puzzling experiments in the triangular lattice antiferromagnet Na$_2$BaCo(PO$_4$)$_2$, 
we propose an interpretation from 
the precursory U(1) DSL within the framework of a mean-field theory. This mean-field theory synthesizes
both the supersolid antiferromagnetic orders and the fractionalized excitations in the DSL. This mixed type 
of mean-field theory has been invoked for the proposed AF$^{\ast}$ state in the high-temperature cuprates,
where the AF$^{\ast}$ state is a $\mathbb{Z}_2$ topologically ordered state with the fermion bilinear antiferromagnetic 
order on the square lattice. 
The ``$\ast$'' in AF$^{\ast}$ refers to the $\mathbb{Z}_2$ topological order with deconfinement and fractionalization 
in the context of Ref.~\onlinecite{lannert2003inelastic}. 
At the level of the mean-field analysis, the treatment and the spirit are quite similar. 
Our analysis in this work will be restricted to the mean-field theory. 
We consider the $\pi$-flux U(1) DSL on the triangular lattice with the supersolid antiferromagnetic order. 
At the mean-field level, we are able to reproduce the continuous excitations around the M and K points 
that were observed by the experiments. 
In addition, as the predicting power of the mean-field analysis, we further predict that, the satellite 
peaks of the dynamic spin structure factor appear at the midpoints between the $\Gamma$ points and the K points.


The remaining parts of the paper are organized as follows. 
In Sec.~\ref{sec2}, we introduce the model Hamiltonian and set up the mean-field theory for both supersolid order
and the spinons. 
In Sec.~\ref{sec3}, we analyze the mean-field theory for the spinons and the supersolid order.
In Sec.~\ref{sec4}, the spectroscopic properties for the spin dynamics are computed from the mean-field theory. 
Finally, in Sec.~\ref{sec5}, we discuss various aspects associated with the triangular lattice easy-axis quantum magnets.
In the appendices, various details of the calculation from the main text are explained.


\section{Model Hamiltonian and mean-field setup}
\label{sec2}

The triangular lattice antiferromagnet Na$_2$BaCo(PO$_4$)$_2$ is modeled   
as the spin-1/2 XXZ model with~\cite{zhong2019strong,NBCP_magnetization}, 
\begin{equation}
H_\text{XXZ} = \sum_{\langle i j \rangle}  J_{\perp} (S_i^x  S_j^x + S_i^y S_j^y) + J_{z} S_i^z S_j^z - B\sum_i S_i^z, 
\label{XXZ}
\end{equation}
where ${J_{\perp}, J_{z} > 0}$ are both antiferromagnetic and the last term 
is the Zeeman effect from an out-of-plane magnetic field $B$. 
With an easy-axis anisotropy ${0<J_{\perp}/J_{z}<1}$, one finds that 
the magnetic phase undergoes through the Y-shape, UUD, 
and V-shape spin configurations with a ${\sqrt{3}\times\sqrt{3}}$ 
three-sublattice structure with increasing fields~\cite{yamamoto2014quantum}, 
and eventually becomes a fully polarized state above the saturation 
field ${B_s={3}J_{\perp}/2+3J_z}$.

The spin orientation on each sublattice is depicted for different ordered phases
in Fig.~\ref{fig: fig1}. Throughout this work, we focus on the zero-field Y-shape state. 
Because the state breaks the U(1) symmetry and the lattice transition at the same time,
this state is also known as the spin supersolid~\cite{wessel2005supersolid,melko2005supersolid}. 
The spin-wave study based on the spin supersolid cannot produce the low-energy continuous 
magnetic excitations near the $\mathbf{M}$ points, indicating the insufficiency of the spin wave theory 
for the spin dynamics. 


\begin{figure}
\includegraphics[width=0.48\textwidth]{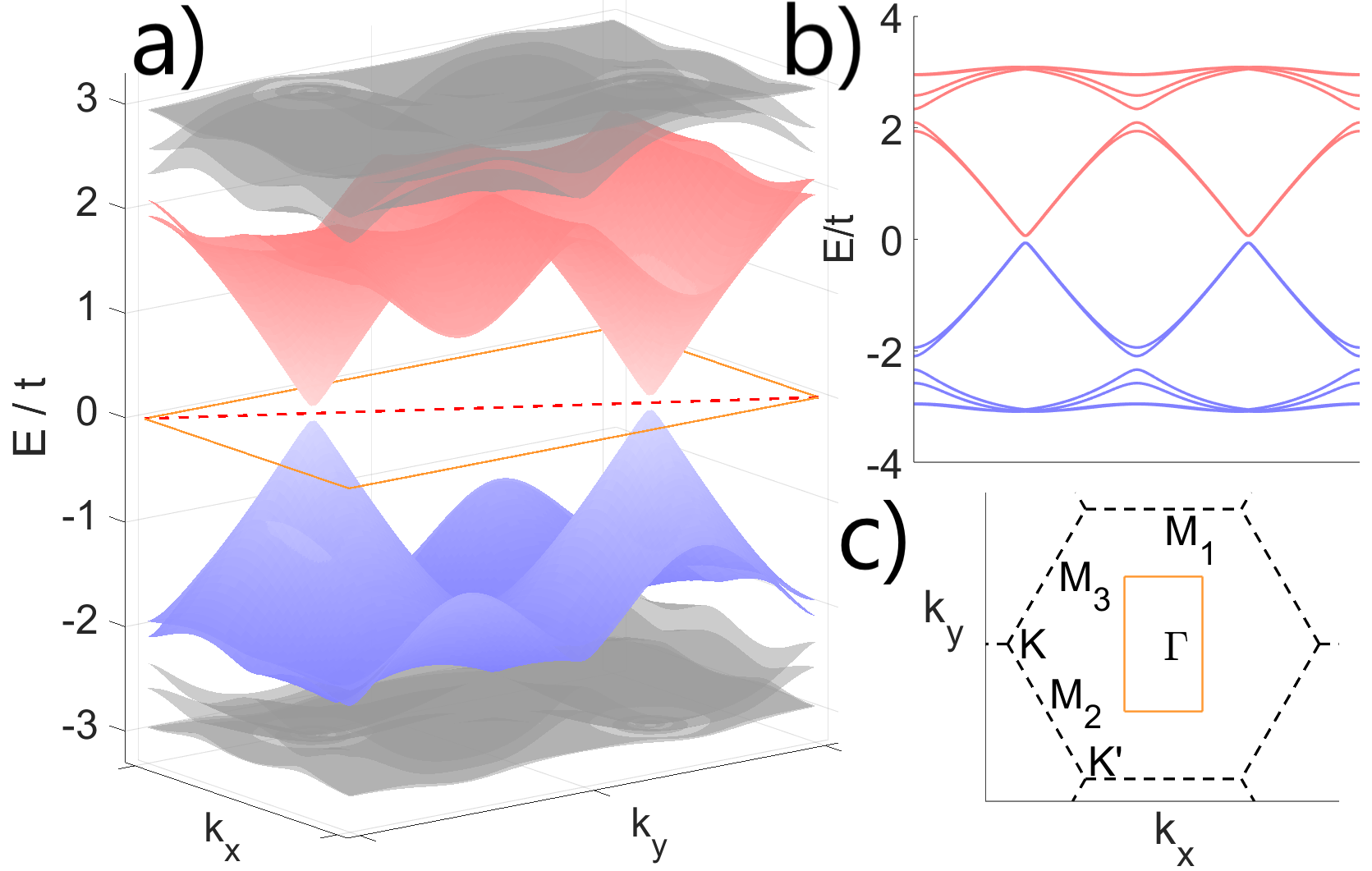}
\caption{(a) The spinon bands in MBZ with ${\tilde{J}_z = 4.394t}$, $\tilde{J}_\perp = 4.102t$, ${\mu=0}$. 
The weakly gapped Dirac cones are relevant for the low-energy INS signals. 
The bands below (above) the Fermi surface are in blue (red). 
The orange rectangle indicates the MBZ. 
(b) The spinon dispersion along the red dashed line in (a).
(c) The hexagonal lattice Brillouin zone with high symmetry points. 
The orange rectangle is the MBZ.}
\label{fig: fig2}
\end{figure}



We aim to find a phenomenological understanding of the zero-field magnetic state 
that contains the supersolid order and at the same time provides the continuous excitations 
near both the K and M points. Based on the recent progress about the U(1) DSL and 
the competing orders~\cite{DSLMotherState}, we thus consider the possibility of the 
precursory DSL for Na$_2$BaCo(PO$_4$)$_2$ and the related easy-axis XXZ model.  
To describe this exotic scenario, we first use the Abrikosov fermionic spinon representation 
for the spin operator with 
\begin{equation}
{{\boldsymbol S}_i = \frac{1}{2} f^\dagger_{i\alpha} {\boldsymbol{\sigma}}_{\alpha\beta} f^{}_{i\beta}},
\end{equation}
and the Hilbert space constraint 
\begin{eqnarray}
{\sum_{\alpha} f_{i\alpha}^\dagger  f_{i\alpha}^{}=1}, 
\end{eqnarray}
where $f_{i\alpha}^{\dagger}$ $(f_{i\alpha}^{} )$ creates (annihilates) a fermionic spinon 
at the site $i$ with the spin $\alpha$. The supersolid antiferromagnetic order is captured 
by the fermionic spinon bilinear in $\langle {\boldsymbol S}_i\rangle$.  
We use the Weiss mean-field theory to incorporate the supersolid order 
and rely on the slave-fermion mean-field theory to capture the U(1) DSL. 
For the first one, one expresses $S^{\mu}_i S^{\mu}_j$ as 
$\langle S^{\mu}_i \rangle S^{\mu}_j $ where $\langle S^{\mu}_i \rangle$ captures the supersolid antiferromagnetic order. 
The continuous spectra in the dynamic spin structure factor 
would be captured by the particle-hole continuum of the fermionic spinons of the DSL at least 
at the mean-field level. It is known that, the $\pi$-flux DSL naturally gives the continuous excitations 
near the M point~\cite{shen2016evidence}.


Combining these two points, we propose the following mean-field Hamiltonian
for ${B=0}$,
\begin{equation}
\begin{split}
    H_\text{MF} &= \sum_{\langle i j \rangle\alpha}  t_{ij}^{} f_{i\alpha}^\dagger f_{j\alpha}^{} - \mu\sum_{i\alpha}f_{i\alpha}^\dagger f_{i\alpha}^{} \\
    &+\sum_{\langle i j \rangle} \tilde{J}_{\perp} \langle S_i^x \rangle S_j^x + \tilde{J}_{\perp} \langle S_i^y \rangle S_j^y + \tilde{J}_{z} \langle S_i^z \rangle S_j^z,
\end{split}
\label{H}
\end{equation}
where we fix the gauge with ${t_{ij} = \pm t}$ (see Fig.~\ref{fig: fig1}(b))
such that a $\pi$-flux through each unit cell is presented, and we have used 
$\tilde{J}_{\perp}$ and $\tilde{J}_{z}$ to distinguish them from the original bare couplings. 
The chemical potential $\mu$ is introduced to enforce the fermion number occupation 
such that 1/2-filling is obtained on average. 
The spinon hopping parameter $t$ can be regarded as the energy unit.
While the chemical potential $\mu$ can be fixed by the fermion filling,
the parameters $\tilde{J}_{\perp}$ and $\tilde{J}_{z}$ can be determined self-consistently
by making sure that the mean-field theory yields the given antiferromagnetic supersolid orders. 

The procedure is briefly described here. From the mean-field Hamiltonian, one
can obtain the spinon mean-field ground state by filling the fermionic spinons up to the spinon Fermi energy ($E_{\text F}$), 
and the ground state is given as 
\begin{eqnarray}
| \psi_0^{\text{MF}} \rangle = \prod_{\epsilon_j < E_{\text F}} f^\dagger_j | 0\rangle ,
\end{eqnarray}
where $\epsilon_j $ refers to the spinon energy for the $j$-th spinon. In reality, one should label the spinon 
with the crystal momentum, the band/sublattice indices, and the spin flavor. One can further perform
the Gutzwiller projection to further improve the wavefunction. Here, we are content with the mean-field treatment 
and aim for a phenomenological description. The magnetic order is then computed as
\begin{eqnarray}
&& \langle S^x_i \rangle = \langle  \psi_0^{\text{MF}}| \hat{S}_i^x    | \psi_0^{\text{MF}} \rangle , \\
&& \langle S^y_i \rangle = \langle  \psi_0^{\text{MF}}| \hat{S}_i^y    | \psi_0^{\text{MF}} \rangle , \\
&& \langle S^z_i \rangle = \langle  \psi_0^{\text{MF}}| \hat{S}_i^z    | \psi_0^{\text{MF}} \rangle ,
\end{eqnarray}
where the left are the input order parameters, and the right are the expectation values of the corresponding spin 
operators with respect to the mean-field ground state. 
The operator hat symbol is introduced explicitly to be distinguished from the left order parameters. 
With the global U(1) symmetry, the above equations can 
be reduced to two equations with one for the in-plane component and the other for the $z$ component. 
The two parameters $\tilde{J}_{\perp}$ and $\tilde{J}_{z}$ can thus be determined. 
Thus, at the mean-field level $H_{\text{MF}}$
describes a $\pi$-flux U(1) DSL with the antiferromagnetic supersolid order, i.e. 
the precursory DSL in the terminology of Sec.~\ref{sec1}.


The original XXZ spin model can be obtained from $H_{\text{MF}}$ by imposing the 
local Hilbert space constraint of having one spinon per site that 
leads to precisely the Hilbert space of the microscopic spin model. This can be 
effectively achieved by introducing a strong onsite Hubbard-$U$-like interaction. 
The spin model can be effectively recovered by doing a second-order 
perturbation of the spinon hopping~\cite{PhysRevB.38.745}. 
The resulting model is of the XXZ form 
with ${J_{\perp} = \tilde{J}_{\perp} +4t^2/U}$ and 
${J_{z} = \tilde{J}_{z} +4t^2/U}$.

\section{Mean-field theory for the spinons and the supersolid order}
\label{sec3}

In the above mean-field setup, we have schematically separated the spin into ``two parts''.
One part is responsible for the antiferromagnetic supersolid order,
and the other part is treated as the spinon bilinears.  Each separately would not able to 
sufficiently capture the continuous excitations that were observed experimentally. We need these
two ingredients together to modify the spin dynamics. 
It is illuminating to write down the spin correlation 
\begin{equation}
    \begin{split}
        \langle \mathbf{S}_i \cdot \mathbf{S}_j \rangle & = 
        \frac{1}{4} \sigma_{\alpha\beta}^\mu \sigma_{\gamma\delta}^\mu 
        \langle f_{i\alpha}^\dagger f_{i\beta} f_{j\gamma}^\dagger f_{j\delta} \rangle \\
        & =  \langle \mathbf{S}_i \rangle \cdot \langle \mathbf{S}_j \rangle 
        - \frac{1}{4} \sigma_{\alpha\beta}^\mu \sigma_{\gamma\delta}^\mu \langle f_{i\alpha}^\dagger f_{j\delta}^{} \rangle
         \langle f_{j\gamma}^\dagger f_{i\beta}^{} \rangle
    \end{split} ,
    \label{correlation}
\end{equation}
where the expectation is taken with respect to the precursory DSL 
that is described by the mean-field theory in the previous section.
Here the first term gives the spin supersolid order, 
and the second term has the spin correlation from the spinons of the DSL.  

In fact, separately computing the $z$-component and $xy$-component spin correlations for 
the nearest neighbors like Eq.~\eqref{correlation}, one could obtain a variational ground state energy $E_{\text{var}}$. 
One could optimize the variational energy for the XXZ model 
against the two variational parameters $\tilde{J}_z$ and $\tilde{J}_{\perp}$. 
The procedure of this approach is slightly different from what has been described in the previous section. 
Here, one does not need the given order parameters that are obtained from the experiments
or the numerical calculations of the XXZ model.

\subsection{Spinon bands from mean-field theory}

To study this precursory DSL, we numerically evaluate the spinon band structures 
of the mean-field Hamiltonian Eq.~(\ref{H}) with the supersolid \textit{ansatz}. 
As the system has a U(1) degeneracy along the $z$-axis, without loss of generality, 
we choose a specific orientation such that 
${\langle S_i^y \rangle = 0}$ (see Fig.~\ref{fig: fig1}). 
Now there remain three degrees of freedom to the \textit{ansatz}, 
which we denote as $\langle S_0^z \rangle$ (the $z$ component of the blue sublattice), 
$\langle S_2^z \rangle$ and $\langle S_2^x \rangle$ (the $z, x$ components of the green sublattice). 
We obtain the following mean-field Hamiltonian, 
\begin{eqnarray}
     H_\text{MF}&=&\sum_{\langle i j \rangle\alpha}  t_{ij} (f_{i\alpha}^\dagger f_{j\alpha}^{} + h. c.)
    -\mu\sum_{i\alpha}f_{i\alpha}^\dagger f_{i\alpha}^{}
\nonumber    \\
    &-&\tilde{J}_\perp \sum_{\langle i j \rangle}
    \left[\frac{\langle S_2^x \rangle}{\sqrt{3}} \sin{({\bf K} \cdot {\bf r}_i)}\right]  
    \sum_{\alpha\beta} f_{j\alpha}^\dagger \sigma^x_{\alpha\beta} f^{}_{j\beta}
    \nonumber \\
&+&\tilde{J}_{z}\sum_{\langle i j \rangle }
 \left[\frac{\langle S_0^z \rangle + 2 \langle S_2^z \rangle}{6} +
 \frac{\langle S_0^z \rangle
  - \langle S_2^z \rangle}{3}\cos(\mathbf{K}\cdot\mathbf{r}_i) \right]
  \nonumber \\
& \times & \sum_{\alpha\beta} f_{j\alpha}^\dagger \sigma^z_{\alpha\beta} f_{j\beta}^{},
 \label{MF}
\end{eqnarray}
where $\langle S_i^{\mu} \rangle$ is extracted from the experiments~\cite{sheng2022two}. More details can be found in Appendix \ref{A}.
The Y-shape state is for the supersolid, where $(\mathbf{M}_0^z, \mathbf{M}_2^z, \mathbf{M}_2^x) = (-0.54, 0.27, 0.173)\mu_\text{B}/\text{Co}$. 
To convert this $\mathbf{M}$ configuration into our spin order, we notice that the in-plane saturation 
moment is ${\mathbf{M}_S^x \approx 2.2\mu_\text{B}/\text{Co}}$, and the out of plane saturation 
moment is ${\mathbf{M}_S^z \approx 2.5\mu_\text{B}/\text{Co}}$~\cite{sheng2022two}. 
Therefore we obtain $\langle S_i \rangle$ according to 
\begin{eqnarray}
(\langle S_0^z \rangle, \langle S_2^z \rangle, \langle S_2^x \rangle) 
&=& \frac{1}{2} (\frac{\mathbf{M}_0^z}{\mathbf{M}_S^z}, \frac{\mathbf{M}_2^z}{\mathbf{M}_S^z}, \frac{\mathbf{M}_2^x}{\mathbf{M}_S^x}) 
\nonumber \\
&=& (-0.108, 0.054, 0.0393).
\end{eqnarray}
Here we do not plan to be quantitative. As it is a spinon mean-field theory without Gutzwiller projection, 
quantitatively reproducing experiments is impractical.  But to be specific, 
we use the above order parameters in the calculation.  

\begin{figure}[t]
\includegraphics[width=0.48\textwidth]{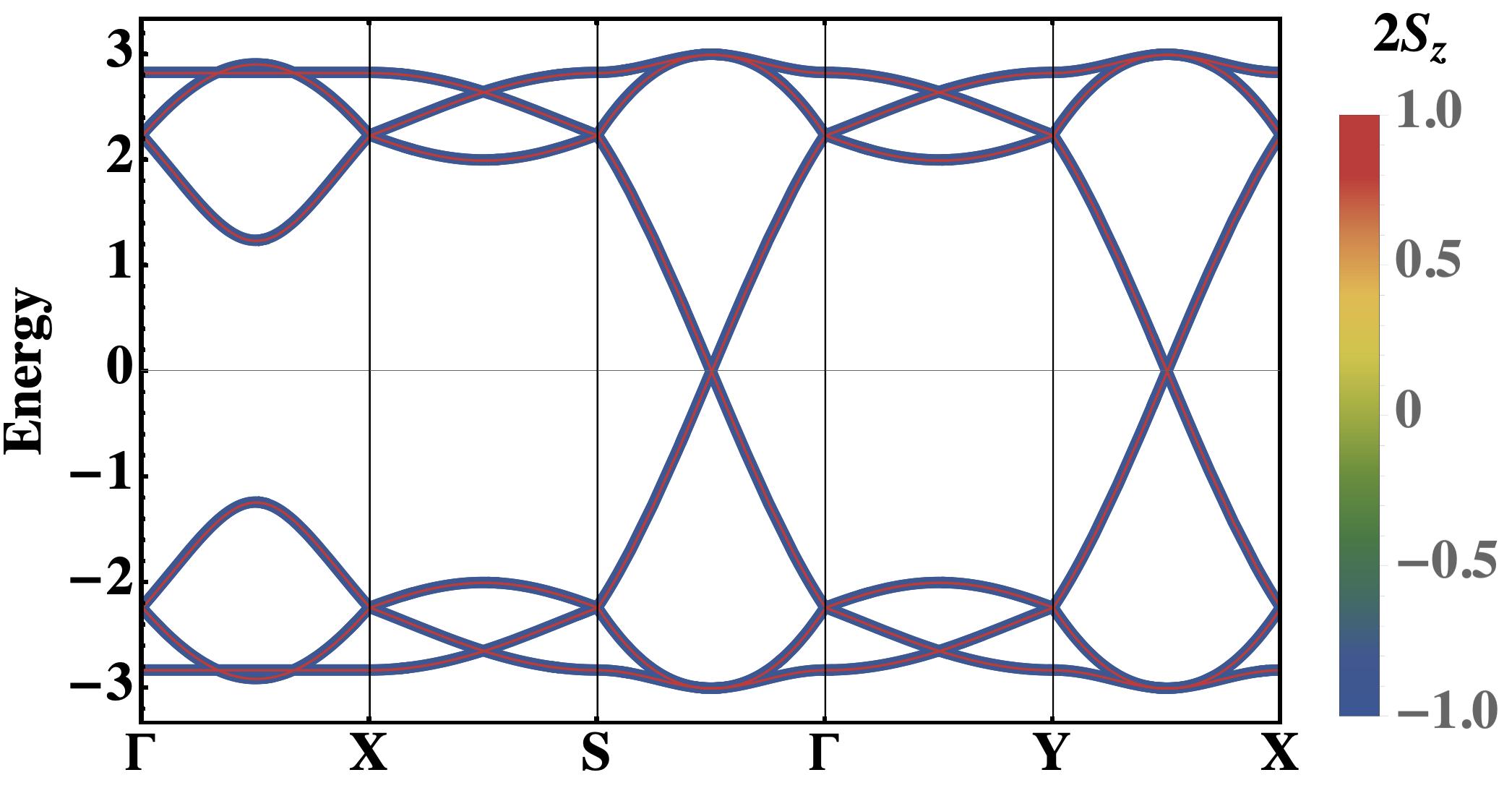}
\caption{The band dispersions for $\tilde{J}_z=0$ and $\tilde{J}_{\perp}=0$. The energy is in the unit of $t$. 
The up-spin spinon bands in red is degenerate with the down-spin spinon bands in blue, and the up-spin and down-spin bands 
are degenerate so the color cannot distinguish them.
The Dirac cones are explicitly shown. }
\label{fig: fig3}
\end{figure}

The spinon hopping can be deduced by comparing the Dirac velocity 
and/or the bandwidth of our INS calculation against the experimental ones~\cite{sheng2024continuum}, 
and we here choose ${t \approx 0.0338 \text{ meV}}$. 
The effective exchange $\tilde{J}$ is obtained self-consistently. 
Namely, we vary $\tilde{J}_\perp$ and $\tilde{J}_z$ to make  
$\langle  \hat{S}^x   \rangle$ and $\langle  \hat{S}^z  \rangle$ to 
be consistent with the input experimental ones. 
 Since ${\langle S_0^z \rangle = -2 \langle S_2^z \rangle}$ 
always holds true (for both experiment and our evaluation), 
there are only two degrees of freedom to vary. 
The result is ${\tilde{J}_z = 4.394t, \tilde{J}_\perp = 4.102t}$.

The self-consistent spinon bands are obtained and shown in Fig.~\ref{fig: fig2}.  
The magnetic unit cell is six times larger than the lattice primitive cell 
due to the supersolid order and the $\pi$-flux background. 
The supersolid order triples the unit cell, and the $\pi$ further doubles the unit cell. 
Thus, with the spin quantum number, 
there are 12 spinon bands in the magnetic Brillouin zone (MBZ) [the orange box 
in Fig.~\ref{fig: fig2}(c)], 
which is 1/6 of the lattice Brillouin zone. In the MBZ, four Dirac cones (2 valleys + 2 spins) 
are separated by a momentum around $\mathbf{M}$. 
In Fig.~\ref{fig: fig2}(b), we further find the Dirac cones 
open up a small gap ${\sim \tilde{J_x} \langle S^x \rangle}$. 
In the next subsection, we separate the Y-shape order into the Ising order 
along the $z$-axis and the planar order in xy-plane, and find that this small gap is opened by 
spin-up/down mixing due to the supersolid order. Since the time reversal symmetry is broken, 
we have made sure that the filled spinon bands have a vanishing Chern number in Appendix \ref{B}, and thus 
there is no Chern-Simons field description or edge mode like the chiral spin liquid.

\begin{figure}[b]
\includegraphics[width=0.48\textwidth]{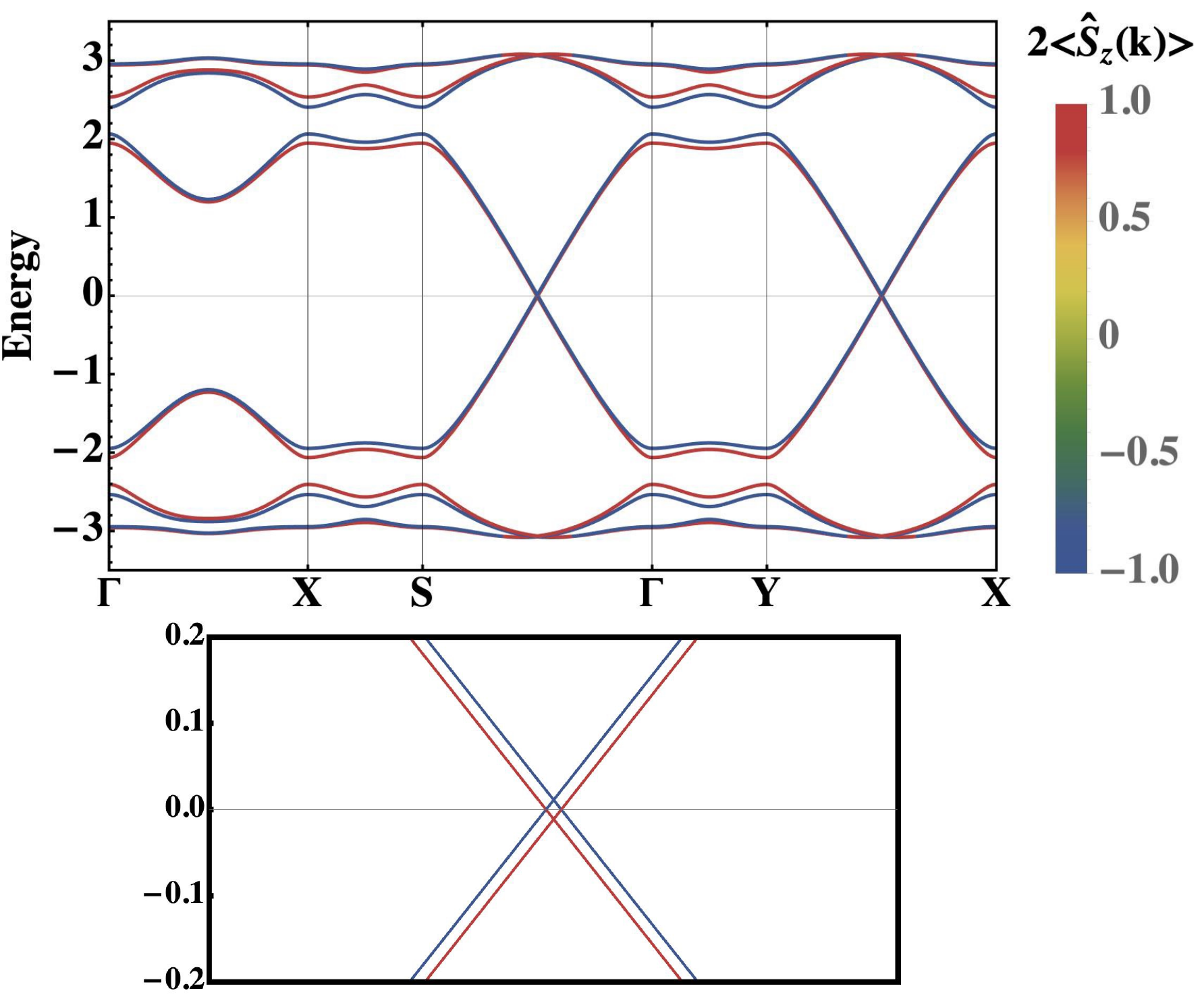}
\caption{The upper figure is the spinon band dispersions for ${\tilde{J}_z |\langle S_0^z\rangle|=0.5t }$ 
and ${ \tilde{J}_\perp\langle S_2^x\rangle=0}$. 
Energy is in the unit of $t$. The red (blue) Dirac cones formed by up-spin (down-spin) bands do not open the gap.
The lower figure is the spinon dispersion near the Fermi level where the gapless feature persists.}
\label{fig: fig4}
\end{figure}

\subsection{Effects of the supersolid order}

In the absence of the spin supersolid order $\langle S_0^z\rangle=\langle S_2^z\rangle=\langle S_2^x\rangle=0$, 
the mean-field Hamiltonian in the previous subsection gives rise to the gapless Dirac cones at the spinon Fermi level. 
There are two Dirac cones for each spin in the MBZ as shown in Fig.~\ref{fig: fig3}. 
The spinon bands are colored by ${\langle\hat{S}_z(\mathbf{k})\rangle=\langle\psi_{n,\mathbf{k}}|\hat{S}_z|\psi_{n,\mathbf{k}}\rangle}$,
where $|\psi_{n,\mathbf{k}}\rangle$ is the wavefunction for the $n$-th spinon band, 
$\hat{S}_z$ is the $z$-component of the spin operator and can be expressed as 
$\mathcal{I}_{6\times6}\otimes\frac{1}{2}\sigma_z$ in the basis of $X_\mathbf{k}$.

To explore the properties of the Dirac spinons, 
in a usual treatment, one analyzes the symmetry properties 
of various mass gap terms for the Dirac cones and establishes the connection 
between the mass gaps with the physical 
spin observables. Here we already have the magnetic orders 
of the system in the beginning. We can then directly compute the evolution of the spinon spectrum
in the presence of the spin supersolid. 
With the supersolid order, we have ${\langle S_0^z\rangle+2\langle S_2^z\rangle=0}$ in Eq.~(\ref{MF}).

\begin{figure}
{\includegraphics[width=0.48\textwidth]{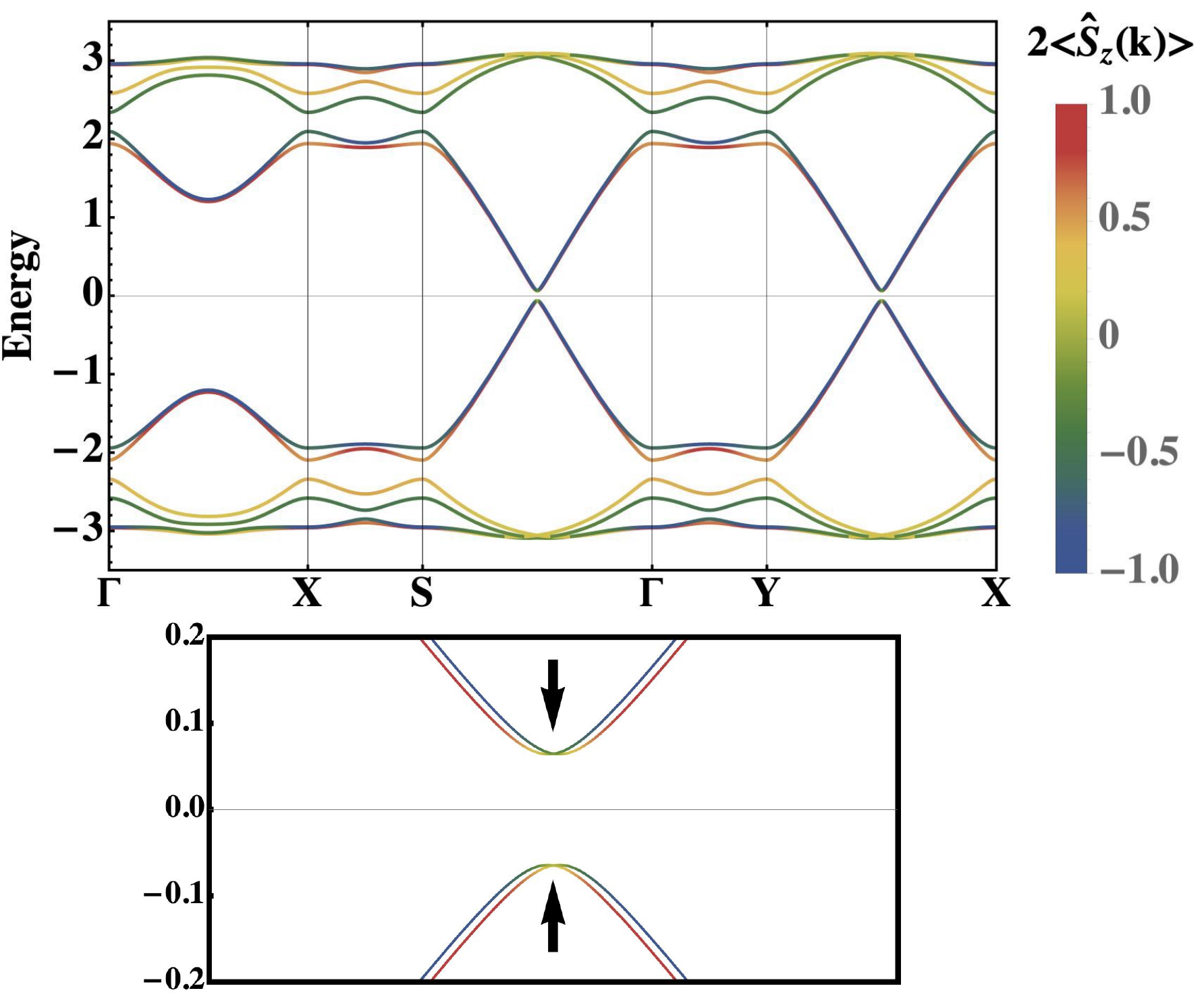}}
\caption{The upper figure is the spinon band dispersion with the same parameters 
that $(\langle S_0^z\rangle,\ \langle S_2^z\rangle,\ \langle S_2^x\rangle)=(-0.108,\ 0.054,\ 0.0393)$ 
and $(\tilde{J}_z,\ \tilde{J}_\perp)=(4.394,4.102)t$. Energy is in the unit of $t$.
The lower figure is the spinon dispersion near the Fermi level where the yellow 
and green bands clearly show that the gaps resulting from the spin-mixing.}
\label{fig: fig5}
\end{figure}

The supersolid order carries both the Ising antiferromagnetic order in $z$ and the 
in-plane $xy$ antiferromagnetic order,
where the former is often referred to as the density wave order and the latter 
is referred to as the superfluid order in the hardcore boson language for the spins. 
Due to the presence of two such orders, it is a bit illuminating to separately study the 
impact on the gapless Dirac cones one by one. 
We begin with the Ising order by setting $\langle S^x \rangle =0$. 
As it is shown in Fig.~\ref{fig: fig4}, the degeneracy
in Fig.~\ref{fig: fig3} is lifted by  $\tilde{J}_z(\langle S_0^z\rangle-\langle S_2^z\rangle)\neq 0$. 
Since the spin-$z$ is still a good quantum number to label the spinon bands, 
the two cones from different spin sectors 
do not mix with each other, and
the Dirac cones at the Fermi level remain gapless 
except for the breaking of the original four-fold degeneracy (see Fig.~\ref{fig: fig4}). 
In the presence of both Ising and in-plane orders for the spin supersolid,
the $\tilde{J}_\perp\langle S_2^x\rangle$ term does not commute with $\hat{S}_z$, 
and this in-plane order immediately creates the gap for the Dirac spinons at the Fermi level by
the spin mixing as shown in Fig.~\ref{fig: fig5}.

\section{Spin dynamics}
\label{sec4}

To connect our results with the inelastic neutron scattering experiments, we further calculate the spinon continuum 
via the transverse dynamic spin structure factor, that is proportional to the inelastic 
neutron spectrum differential cross-section, as 
\begin{eqnarray}
    {\mathcal S}^{-+}(\mathbf{p}, E) 
        & &= \frac{1}{N} \sum_{i, j} e^{i \mathbf{p} \cdot (\mathbf{r}_i - \mathbf{r}_j)} \int e^{i E t} \langle S_i^-(t) \cdot S_j^+(0) \rangle  \dd t 
        \nonumber \\
        & &= \sum_n \delta(E - [E_n(\mathbf{p}) - E_0]) \left| \mel{n}{S_\mathbf{p}^+}{\Omega} \right| ^2,
 \label{S}
\end{eqnarray}
where $\ket{\Omega}$ ($\ket{n}$) and $E_0$ ($E_n$) refer to the ground (excited) state and its energy, respectively. 
At the mean-field level, ${\ket{\Omega}}$ is to simply fill the spinon bands below the Fermi energy, 
and ${ S_{\bf p}^+ = \sum_{\bf k} f_{{\bf k} - {\bf p} \uparrow}^\dagger f_{{\bf k} \downarrow} }$ 
excites one spinon particle-hole pair across the Fermi level. 
The summation over $n$ includes all such excited spin-1 pairs. 
Since the neutron scattering events involve pairs of spinons, the energy and momentum 
transfer of one neutron are conserved by the total energy and total momentum of two spinons. 
This leads to a continuous spectrum in the INS measurements, 
that represents the fractionalized excitations~\cite{shen2016evidence,han2012fractionalized,paddison2017continuous} 
and differs from the magnon excitation with integer spins.

\begin{figure}
\centering
\includegraphics[width=0.48\textwidth]{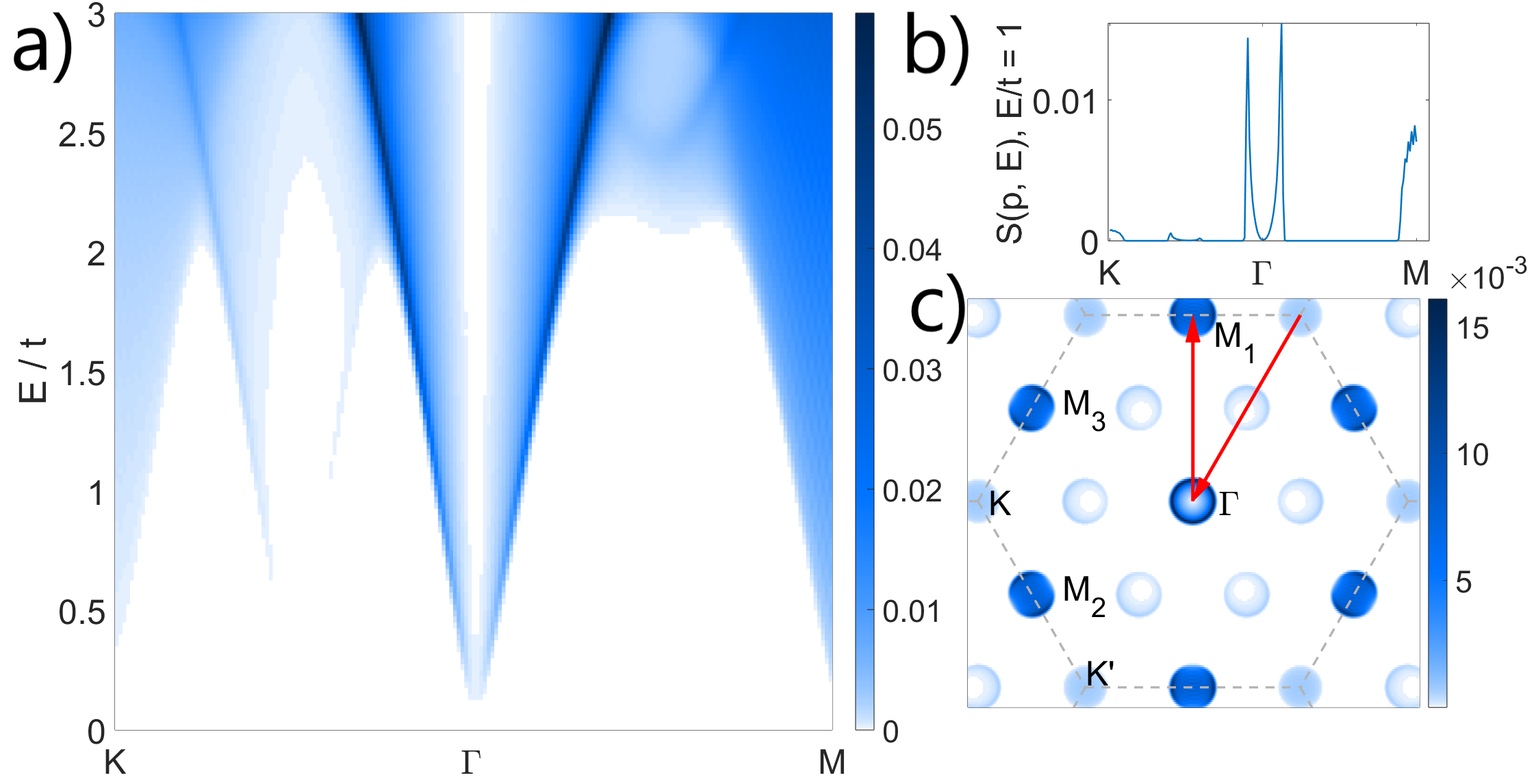}
\caption{ 
(a) The INS spectrum along the high-symmetry path 
with the same parameters in Fig.~\ref{fig: fig2}. 
In the low-energy regime of the INS spectrum, there are two strong signals 
at ${\bf \Gamma}$ and $\mathbf{M}$, and lighter ones at $\mathbf{K}$ and $\mathbf{K}/2$. 
(b) The constant energy slice of (a) at ${E/t = 1}$ and is marked by the red arrow in (a).
(c) 
The momentum dependence of ${\mathcal S}^{-+}(\mathbf{p}, E)$ 
at ${E/t = 1}$.  
 }
\label{fig: fig6}
\end{figure}

In Fig.~\ref{fig: fig6}, we show the continuum along the high symmetric momentum path. 
There are two additional features that arise here. 
First, the expected signals appear at $\mathbf{M}$ in addition to  
${\bf K}$ at low energies, 
which agrees with the experimental results.
Second, the model predicts the satellite peaks at the middle point from ${\bf \Gamma}$ and ${\bf K}$. 
Fig.~\ref{fig: fig6} shows the INS signals in the whole Brillouin zone 
with a fixed energy slice, including the ones around $\{\mathbf{\Gamma},\mathbf{M}_{1, 2, 3}\}$ 
and $\{\mathbf{K}, \mathbf{K}', \mathbf{K}/2,\mathbf{K}'/2\}$. 
These main features persist even with a weak external magnetic field as shown in Appendix \ref{C}.

The position of these signals is a direct consequence of the coexistence of the DSL and the supersolid order. 
At low energies, the spinon particle-hole pairs can be created only across the Fermi energy 
among those gapped Dirac cones. 
Therefore, the allowed momentum transfer $\mathbf{p}$'s in Eq.~(\ref{S}) are about the separation of Dirac cones.
As shown in Fig.~\ref{fig: fig6}, the distribution of the Dirac cones 
indeed matches the pattern of the INS signal in the momentum space. 
The spectral continuum at $\{\mathbf{K},\mathbf{K}',\mathbf{K}/2,\mathbf{K}'/2\}$ is 
then a consequence due to the combination of the Dirac spinons and the supersolid order 
that has an ordering wavevector at ${\mathbf{K}}, \mathbf{K}'$. 
Since our theory is based on the free spinons and the Hilbert space constraint is not 
imposed on each site, the intra-Dirac-cone processes that contribute to the $\Gamma$ point
should not be favored by the antiferromagnetic spin interaction, and would in principle be suppressed. 
This can be shown with a random phase approximation to incorporate the spinon interactions~\cite{PhysRevB.96.075105}.
This is the caveat of the free spinon theory~\cite{PhysRevX.9.031026}.

\begin{figure}
\includegraphics[width=0.48 \textwidth]{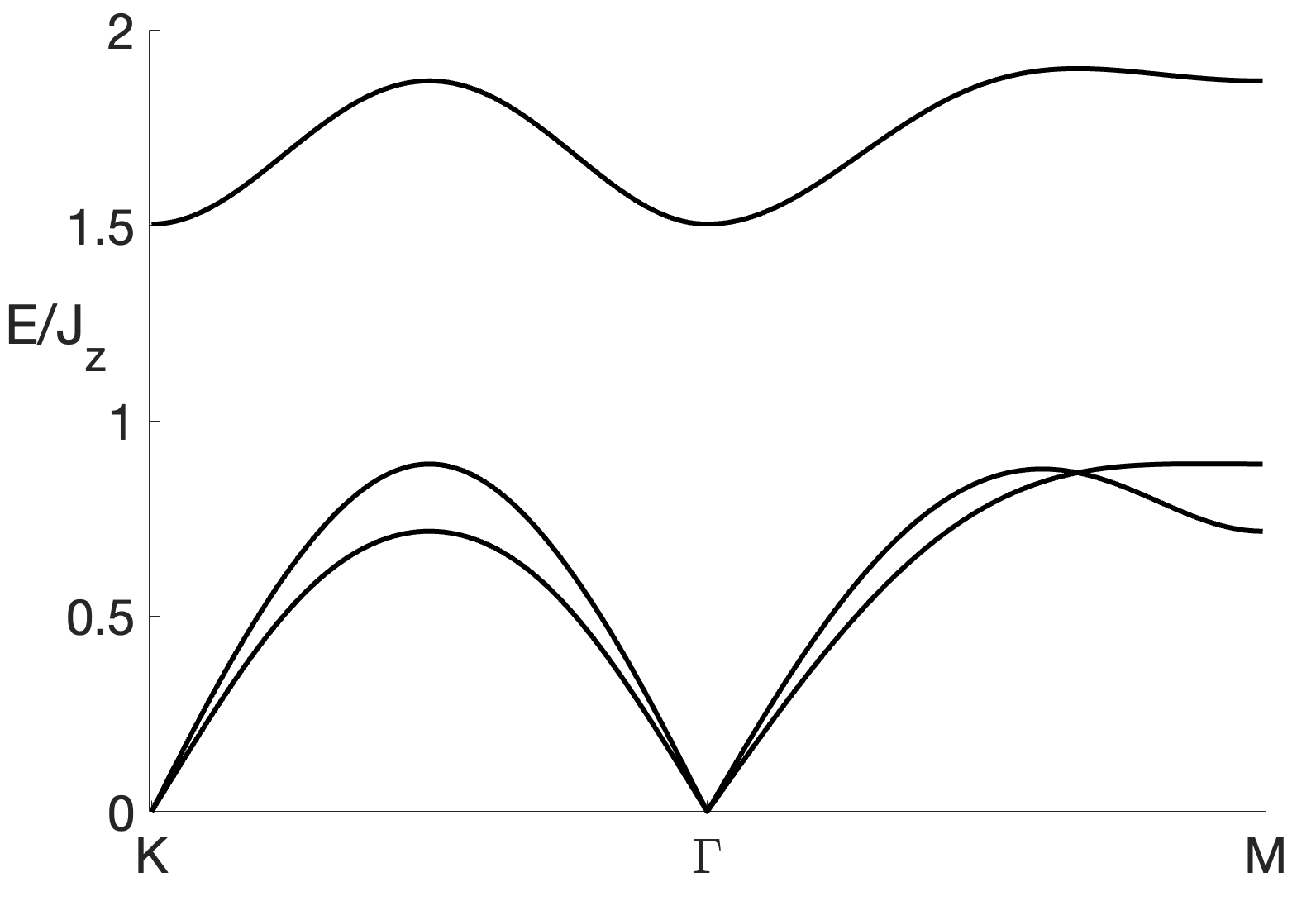}
\caption{The spin-wave excitation from linear spin-wave theory for the supersolid state. 
For the Y-shape configuration, the tilting angle $\theta$ from $\hat{z}$
on the red and green sublattices (see Fig.~\ref{fig: fig1}) 
is determined by minimizing the classical exchange energy with 
${\cos{\theta} = J_z/(J_z + J_{\perp})}$ and ${J_z =1.73J_{\perp}}$. 
Although this classical state is a ferrimagnet with a finite magnetization
and differs slightly from the quantum supersolid,
the spin-wave excitation should be a reasonable qualitative approximation 
of the spin-wave part of the excitations for the quantum supersolid. 
}
\label{fig: fig7}
\end{figure}

Although we have obtained the continuous excitation from spinons, 
the fluctuation of the order parameters can still be important,
and these are spin-wave excitations as magnons. 
These magnons can also contribute to the INS spectrum,
but not usually in terms of continuous excitation. 
The magnons contribute to the INS spectrum 
as a well-defined quasiparticle-like dispersion.
We calculate the spin-wave dispersion with
the linear spin wave theory,
and the results are depicted in Fig.~\ref{fig: fig7}. 
The gapless spectrum is due to the breaking of 
the U(1) symmetry. The spectrum at $\mathbf{K}$ is understood by 
the supersolid order with a wavevector $\mathbf{K}$,
and there is a visible gap at {\bf M} point. 
In reality, the magnetic order is strongly renormalized by the quantum
correction~\cite{sheng2022two}, and the bandwidth of the spin-wave excitation should be 
suppressed compared to the linear spin-wave calculation.  
Taking together with the spinon continuum, 
the actual INS signal should be a combination of both 
spinon continuum and the magnonic excitations~\cite{bose2023proximate,lannert2003inelastic}.

\section{Discussion}
\label{sec5}

There are two major questions that are addressed in this work.
One is whether the continuous excitation in Na$_2$BaCo(PO$_4$)$_2$ arises from 
the spinon pairs of any sort. The other question is whether Na$_2$BaCo(PO$_4$)$_2$ 
has anything to do with the DSL. 
We provide a candidate answer to both questions by proposing a precursory 
DSL that combines the DSL and the supersolidity. 
It turns out that, our treatment is not really new. As we have recently learned, 
a similar approach was actually used by Ref.~\onlinecite{PhysRevLett.104.015301} to 
establish the supersolid spin order, instead of the spin dynamics. 
The progress in our work is to access the spin dynamics with this approach.

At the mean-field level  
we have obtained two-spinon continuum and found concentrated signals 
at $\{\mathbf{M},\mathbf{K}, \mathbf{K}/2\}$ points. 
These results provide a possible explanation for the bizarre INS results in 
Na$_2$BaCo(PO$_4$)$_2$. 
For the first question, phenomenologically two-magnon continuous excitations
could also contribute to the INS continuum~\cite{lovesey1984theory}.
This magnon continuum, however, usually occurs at higher energies instead of the low energy, 
and its spectral intensity is usually weak. The recent work in Ref.~\onlinecite{sheng2024continuum} 
actually calculated the two-magnon contribution and did not find an agreement with 
As for other candidate states, one natural candidate seems to be a $\mathbb{Z}_2$ 
spin liquid with condensed bosonic spinons that generate the supersolid order. 
We find it difficult to reconcile the magnetic structure and the position of the continuous excitations
of the $\mathbb{Z}_2$ spin liquids that were classified with Schwinger bosons~\cite{PhysRevB.74.174423}. 
Phenomenologically, however, it is possible that, if one 
directly introduces the supersolid on top of the $\pi$-flux 
$\mathbb{Z}_2$ spin liquid in Ref.~\onlinecite{PhysRevB.74.174423}, 
one may establish a qualitatively consistent result with the experiment. 
We do not rule out this possibility here.  
Since the model for Na$_2$BaCo(PO$_4$)$_2$ is the easy-axis XXZ model, 
we expect more advanced numerical techniques to be used for the 
spin dynamics. In fact, the $J_1$-$J_2$ Heisenberg model on 
the triangular lattice was numerically found to stabilize the U(1) DSL with a very weak $J_2$. 
The easy-axis XXZ model is more frustrated than the nearest-neighbor Heisenberg model, and is 
likely to be more close to the DSL.

Our free-spinon mean-field theory would certainly be 
renormalized once the Hilbert space constraint is implemented. 
This can be achieved by the renormalized mean-field 
theory or by the Gutzwiller projection~\cite{00018730701627707}. 
Except for the modulation and suppression of the spectral intensities, 
the qualitative results for the spinon continuum along with the proximate spin liquid state should persist,
{so we do not invoke such methods as we do not focus on the precise spectrum calculations in this study}. 
With the supersolid, the precursory DSL 
develops a small gap, and this can potentially cause the confinement 
of the U(1) gauge theory. Since the U(1) DSL is regarded as the {\sl mother}
state for 2D magnets~\cite{DSLMotherState}, one could 
imagine a weak spinon pairing that further stabilizes fractionalization~\cite{PhysRevB.93.165113}. 
Such a state could be equivalent to a $\pi$-flux $\mathbb{Z}_2$ spin liquid
from the Schwinger boson classification~\cite{PhysRevB.74.174423}. 
Nevertheless, as long as the fractionalization is present below the confinement length, 
one would expect continuous excitations at finite energies. 
Moreover, thermal fluctuation suppresses the supersolid above 0.15K,
and our theory expects the system to behave like a gapless DSL. 
In fact, a finite thermal conductivity $\kappa/T$ was observed at low temperatures
above 0.15K~\cite{NBCP_magnetization} and is consistent
with the expectation from DSL~\cite{PhysRevB.62.1270}.  


Beyond Na$_2$BaCo(PO$_4$)$_2$, other Co-based triangular lattice antiferromagnetic materials 
with the easy-axis XXZ model description could exhibit similar physics, i.e. the coexistence
of the magnetic order and the spinon continuum (especially around the $M$ points).
In fact, a recent inelastic neutron scattering measurement in another Co-based triangular 
lattice antiferromagnet K$_2$Co(SeO$_3$)$_2$ also found quite similar structures of continuous excitations,
and this system is clearly in the easy-axis regime of the XXZ model with the supersolid order~\cite{zhu2024continuum}. 
Moreover, this measurement discovered the low-energy satellite peaks at ${\bf K}/2$ and ${\bf K}'/2$
of the continuous excitations, and this is consistent with our theoretical expectation. 
The isostructural material Rb$_2$Co(SeO$_3$)$_2$ is expected to behave quite similarly. 


Ref.~\onlinecite{DSLMotherState} has suggested searching for the precursory DSL 
among 2D frustrated lattices. One related such system that behaves like a DSL 
is the spin liquid candidate PrZnAl$_{11}$O$_{19}$ where the Pr ions form a triangular 
lattice with spin-1/2 local moments~\cite{Pr_based_first, Pr_based_gang}. For these 
triangular lattice antiferromagnets with the non-Kramers doublets, the recent model    
only slightly deviates from the XXZ model by allowing one extra anisotropic interaction~\cite{Liu_2018}. 
Continuous excitations have actually been observed in PrZnAl$_{11}$O$_{19}$~\cite{PhysRevB.109.165143}. 
In addition to the frustrated lattices, the bipartite lattices with frustrated interactions can 
still be promising candidates, as recently proposed~\cite{bose2023proximate,PhysRevB.107.184423,PhysRevB.109.024419}. 
On more theoretical side, 
the fermionized vortex theory was used to describe the DSL on the triangular lattice~\cite{PhysRevLett.95.247203}
and may encounter some issues with imposing time reversal symmetry. Since the time reversal
is broken by the supersolid, it might be interesting to revisit the fermionized vortex theory 
to explore the mass generation and supersolid in the DSL.  
An alternative approach is to directly fermionize the hardcore bosons that represent the spins~\cite{PhysRevLett.87.097203}. 
This approach is more direct, and can be more useful to obtain the variational wavefunction for the ground state. 
Numerically, it would be nice to directly access the DSL with the XXZ $J_1$-$J_2$ model.


\begin{acknowledgments}
We particularly thank Liusuo Wu for sharing his data prior to the submission, and Cenke Xu and Jiawei Mei for discussion. 
This work is supported by NSFC with Grant No.~92065203, MOST of China with Grants No.~2021YFA1400300,  
and by the Collaborative Research Fund of Hong Kong with Grant No. C6009-20G
and C7012-21G, by the Guangdong-Hong Kong Joint
Laboratory of Quantum Matter, the NSFC/RGC JRS
grant with Grant No. N HKU774/21, by the General Research Fund of Hong Kong with Grants No. 17310622
and No. 17303023, and by the Fundamental Research Funds for the Central Universities, Peking University.
\end{acknowledgments}
 
\appendix

\section{Mean-field theory}\label{A}

We choose the lattice vectors as ${\mathbf{a}_1=(a,0) }$ 
and $\mathbf{a}_2=(a/2, \sqrt{3}a /2)$ as shown in Fig.~1 of
the main text. 
Then, the basis vectors for the reciprocal lattice are 
$\mathbf{b}_1=({2\pi}/{a},-{2\pi}/{\sqrt{3}a})$ and ${\mathbf{b}_2=(0,{4\pi}/{\sqrt{3}a})}$. 
The high-symmetry momentum points are then defined as 
$\mathbf{M} = \mathbf{b}_2/2$, $\mathbf{K} =  2\mathbf{b}_1/3+\mathbf{b}_2/3$.
We will stick to this convention in the appendix. 

The physical spin model is a spin-1/2 XXZ model on the triangular lattice
\begin{equation}
    H_{\text{XXZ}} = \sum_{\langle i j \rangle}  J_{\perp} (S_i^x  S_j^x + S_i^y S_j^y) + J_{z} S_i^z S_j^z,
\end{equation}
where $J_{\perp}$ and $J_z$ are the original spin exchange couplings. 
To capture both the supersolid order and the continuous excitations of this XXZ model, 
we implement a mean-field approach that combines the conventional 
Weiss type of mean-field theory for the magnetic order and the 
parton mean-field theory to take care of the exotic excitations. 
In the following, we provide a more detailed description of the mean-field theory 
in the main text.

\subsection{Weiss mean-field channel}

For the Weiss mean-field channel where $\langle S_i^\mu \rangle\neq 0$, 
we use a mean-field \textit{ansatz} that is implied by the existing experiments 
and numerical studies,
\begin{align}
\left\{\begin{array}{lll}
     \langle S_i^x \rangle = \frac{2 \langle S_2^x\rangle\cos\phi}{\sqrt{3}} \sin{(\mathbf{K} \cdot \mathbf{ r}_i)}  ,
     \vspace{2mm}\\
     \langle S_i^y \rangle = \frac{2 \langle S_2^x\rangle\sin\phi}{\sqrt{3}} \sin{(\mathbf{K} \cdot \mathbf{ r}_i)},
     \vspace{2mm}\\
\langle S_i^z \rangle = \frac{\langle S_0^z\rangle+2\langle S_2^z\rangle}{3} + 2\frac{\langle S_0^z\rangle 
                                  - \langle S_2^z\rangle}{3} \cos{(\mathbf{K} \cdot \mathbf{r}_i)},
\end{array}
\right.
\label{SS}
\end{align}
so that a three-sublattice Y-shape supersolid antiferromagnetic ordering is produced as
\begin{align}
\left\{\begin{array}{lll}
\mathbf{S}_{(0,0)}=(0,0,\langle S_0^z\rangle) ,
\vspace{2mm}\\
\mathbf{S}_{\mathbf{a}_1}=(-\langle S_2^x\rangle\cos\phi,-\langle S_2^x\rangle\sin\phi,\langle S_2^z\rangle),
\vspace{2mm}\\
\mathbf{S}_{\mathbf{a}_2}=(\langle S_2^x\rangle\cos\phi,\langle S_2^x\rangle\sin\phi,\langle S_2^z\rangle),
\end{array}\right. 
\end{align}
while $\langle \mathbf{S}_0\rangle$ 
and $\langle \mathbf{S}_2\rangle$ 
refer to the spin orders on the corresponding sublattices 
and can be determined self-consistently
from the mean-field theory. 
Moreover, $\langle S^z_0 \rangle =-2 \langle S^z_1 \rangle$. 
For the convenience, we set $\phi=0$.

We express the spin operator using the Abrikosov spinon representations 
${S_i^\mu=\frac{1}{2}f^\dagger_{i\alpha}\sigma^\mu_{\alpha\beta}f_{i\beta}}$ 
with the constraint ${\sum_\alpha f_{i\alpha}^\dagger f_{i\alpha}^{} = 1}$.
Then, the supersolid's contribution to the mean-field Hamiltonian 
in the main text can be written as
\begin{eqnarray}
    H_\text{SS}&= &-\tilde{J}_\perp\sum_{j,\boldsymbol{\delta}_i}
    \left[\frac{\langle S_2^x\rangle}{\sqrt{3}} \sin{({\bf K} \cdot \mathbf {r}_i)}\right]
    (e^{-i\phi}f_{j\uparrow}^\dagger f_{j\downarrow}^{}+h.c.)
    \nonumber \\
    &+&  \tilde{J}_{z}\sum_{j,\boldsymbol{\delta}_i}
    \left[\frac{\langle S_0^z\rangle+2\langle S_2^z\rangle}{6} + \frac{\langle S_0^z\rangle 
    - \langle S_2^z\rangle}{3} \cos{(\mathbf{K} \cdot \mathbf{r}_i)}\right]
\nonumber \\
&\times& (f_{j\uparrow}^\dagger f_{j\uparrow}^{}-f_{j\downarrow}^\dagger f_{j\downarrow}^{} ),\nonumber \\
 \label{SS_MF}
\end{eqnarray}
where ${\mathbf{r}_i=\mathbf{r}_j+\boldsymbol{\delta}_i}$ 
with $ \boldsymbol{\delta}_i=\{ \pm \mathbf{a}_1,\pm \mathbf{a}_2,\pm{\mathbf{a}_3} \}$
with $ {\mathbf{a}_3} ={\mathbf{a}_2-\mathbf{a}_1}$ 
denoting the six nearest neighbours. 
Following the treatment of the main text, we have introduced the $\tilde{J}_{\perp}$
and $\tilde{J}_{z}$ couplings for the supersolid part of the mean-field theory.  
Moreover, we adopted 
the mean-field decoupling with, 
\begin{eqnarray}
\sum_{\langle i j \rangle}S_i^\alpha S_j^\alpha \rightarrow \sum_{\langle i j \rangle}\langle S_i^\alpha\rangle S_j^\alpha+ S_i^\alpha\langle S_j^\alpha\rangle-\langle S_i^\alpha\rangle\langle S_j^\alpha\rangle.
\end{eqnarray}

\begin{figure}[t]
\includegraphics[width=0.4\textwidth   ]{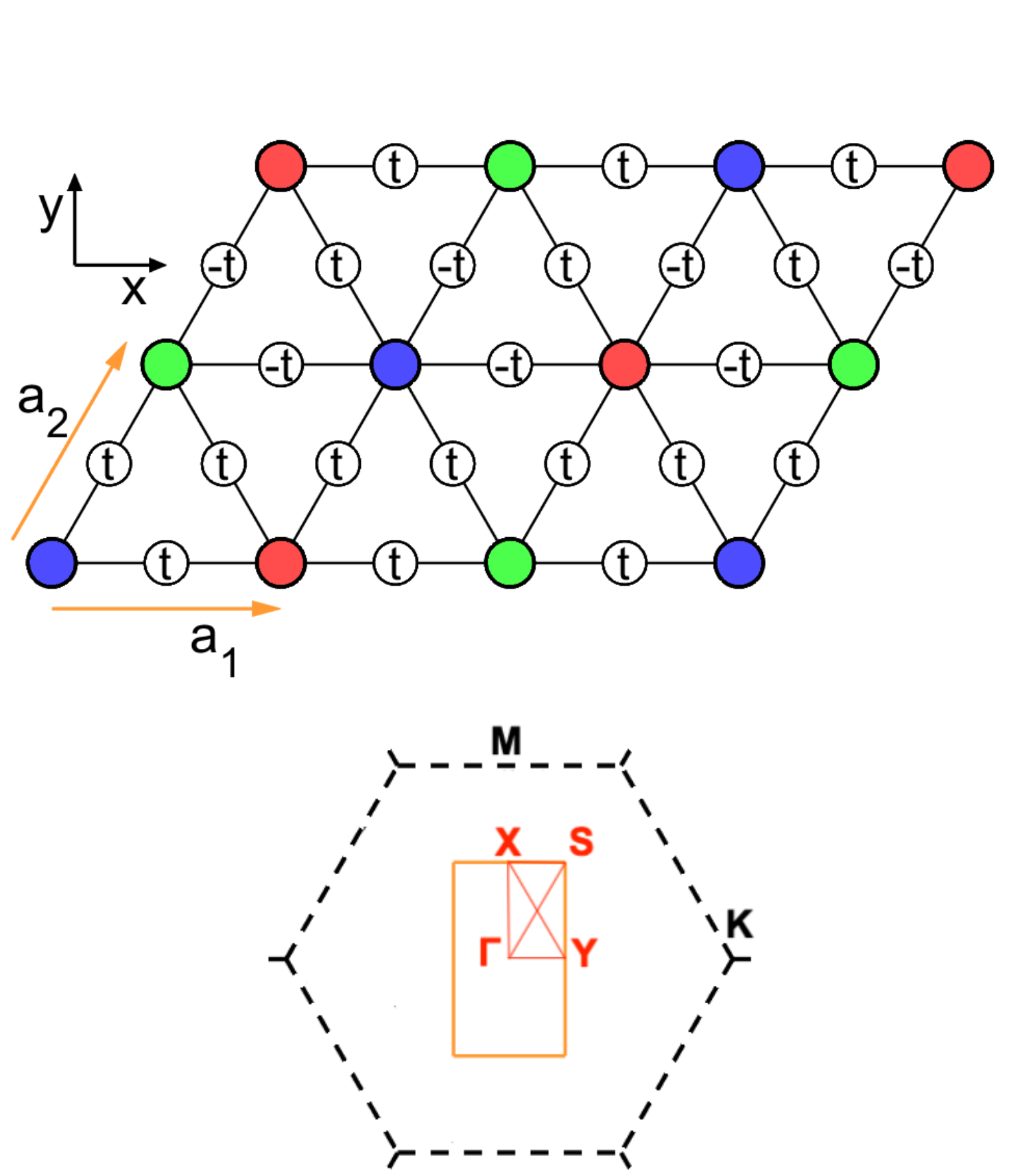} \\
 \caption{The upper figure is the triangular lattice formed by the Co$^{2+}$ ions with the basis vectors ${\mathbf{a}_1=a(1, 0)}$ and $\mathbf{a}_2=a(1/2,\sqrt{3}/2)$. 
 The hopping parameters $t_{ij}$ of the spinons are indicated within the bond with a gauge choice supporting a $\pi$-flux background. 
In the lower figure, the black dashed lines show the lattice Brillouin zone and the orange solid lines show the magnetic Brillouin zone. 
 The corresponding high symmetry points are also denoted.}
\label{fig: fig8}
\end{figure}

With the Fourier transform ${f_{i\sigma}=1/\sqrt{N} \sum_{\mathbf{k}}e^{i\mathbf{k}\cdot\mathbf{r}_i}f_{\mathbf{k}  \sigma} }$, 
where $N$ is the system size and the sum is over the lattice Brillouin zone (BZ),
\begin{widetext}
\begin{eqnarray}
     H_\text{SS}&=&\frac{\sqrt{3}
     \tilde{J}_\perp\langle S_2^x\rangle}{2i}\sum_{ \mathbf{k} \in \text{BZ}}
     \left[ \left(f_{\mathbf{k+K},\uparrow}^\dagger f_{\mathbf{k}\downarrow}^{}-f_{\mathbf{k-K},\uparrow}^\dagger f_{\mathbf{k}\downarrow}^{}\right)+ \left(f_{\mathbf{k+K},\downarrow}^\dagger f_{\mathbf{k}\uparrow}^{}-f_{\mathbf{k-K},\downarrow}^\dagger f_{\mathbf{k}\uparrow}^{}\right)\right]
     \nonumber \\
    && +\tilde{J}_z\sum_{\mathbf{k}\in \text{BZ}}
    \Big[
    (\langle S_0^z\rangle+2\langle S_2^z\rangle)(f_{\mathbf{k}\uparrow}^\dagger f_{\mathbf{k}\uparrow}^{}
    -f_{\mathbf{k}\downarrow}^\dagger f_{\mathbf{k}\downarrow}^{})+\frac{\langle S_2^z\rangle-\langle S_0^z\rangle}{2}(f_{\mathbf{k+K},\uparrow}^\dagger f_{\mathbf{k}\uparrow}^{}+f_{\mathbf{k-K},\uparrow}^\dagger f_{\mathbf{k}\uparrow}^{}
         \nonumber \\
&&    \quad\quad\quad\quad  -f_{\mathbf{k+K},\downarrow}^\dagger f_{\mathbf{k}\downarrow}^{}
    -f_{\mathbf{k-K},\downarrow}^\dagger f_{\mathbf{k}\downarrow}^{} ) \Big].
    \label{SS_MF_k}
 \end{eqnarray}
Here, we used $\sum_{\boldsymbol{\delta}_i} e^{i \mathbf{K}\cdot\boldsymbol{\delta}_i}=-3$.

\subsection{Spinon hopping channel}

For the spinon hopping channel ($\langle f^\dagger_{i} f_{j}\rangle \neq 0$), 
as the neutron scattering data indicate the presence of a percursory Dirac spin liquid (DSL), 
we consider the \textit{ansatz} allowed by symmetries: a U(1) DSL with only a real nearest 
neighbor hopping [See Fig.~\ref{fig: fig8}]. 

With this \textit{ansatz}, the spinon hopping part of the mean-field theory can be expressed as
   \begin{equation}
\begin{split}
    H_t&=t\sum_{i\sigma}\left[\cos\left({\mathbf{M}\cdot\mathbf{r}_i}\right)\left(f_{i\sigma}^\dagger f_{i+\mathbf{a}_1,\sigma}^{}
    +f_{i\sigma}^\dagger f_{i+\mathbf{a}_2,\sigma}^{} \right)
    +f_{i\sigma}^\dagger f_{i+\mathbf{a}_3,\sigma}^{}+ h. c.\right] -\mu\sum_{i\sigma}f_{i\sigma}^\dagger f_{i\sigma}^{}\\
    &=\sum_{\mathbf{k} \in \text{BZ}}\sum_{\sigma}(2t\cos\left({\mathbf{k}\cdot\mathbf{a}_3}\right)-\mu) f_{\mathbf{k}\sigma}^\dagger f_{\mathbf{k}\sigma}^{} \\
    &+\sum_{\mathbf{k} \in \text{BZ}}\sum_{\sigma}\left[t\frac{e^{i\mathbf{k}\cdot\mathbf{a}_1}+e^{i\mathbf{k}\cdot\mathbf{a}_2}}{2}\left(f_{\mathbf{k+M},\sigma}^\dagger f_{\mathbf{k}\sigma}^{}+f_{\mathbf{k-M},\sigma}^\dagger f_{\mathbf{k}\sigma}^{} \right)+t\frac{e^{-i\mathbf{k}\cdot\mathbf{a}_1}+e^{-i\mathbf{k}\cdot\mathbf{a}_2}}{2}\left(f_{\mathbf{k}\sigma}^\dagger f_{\mathbf{k-M},\sigma}^{} +f_{\mathbf{k}\sigma}^\dagger f_{\mathbf{k+M},\sigma}^{} \right)\right]\\
    &=\sum_{\mathbf{k}\in \text{BZ} ,\sigma} (2tA_\mathbf{k}-\mu) f_{\mathbf{k}\sigma}^\dagger f_{\mathbf{k}\sigma}^{}
    +tB_\mathbf{k}f_{\mathbf{k}\sigma}^\dagger f_{\mathbf{k+M},\sigma}^{}+tB^*_\mathbf{k}f_{\mathbf{k+M},\sigma}^\dagger f_{\mathbf{k}\sigma}^{},
    \label{Hopping_MF}
\end{split}
\end{equation} 
\end{widetext}
where ${A_\mathbf{k}=\cos\left({\mathbf{k}\cdot\mathbf{a}_3}\right)},\ 
{B_\mathbf{k}=\cos{(\mathbf{k}\cdot\mathbf{a}_1)}-i\sin{(\mathbf{k}\cdot\mathbf{a}_2)}}$, 
and we have used the fact that $2\mathbf{M}=\mathbf{b}_2$.

Since the U(1) DSL \textit{ansatz} gives rise to a $\pi$-flux background and doubling of the spinon unit cell, 
along with
the presence of the supersolid ordering, the size of the magnetic Brillouin zone (MBZ) is 1/6 as large as the size of 
the lattice Brillouin zone in Fig.~\ref{fig: fig8}. 
In order to obtain the dispersion of the spinon excitations, we need to fold the lattice Brillouin zone (BZ) 
into MBZ and formally convert the sum over $\mathbf{k}\in\text{BZ}$ in Eqs.~\eqref{SS_MF_k} and \eqref{Hopping_MF} into the summation
over $\mathbf{k}'\in\text{MBZ}$ as
\begin{equation}
\sum_{\mathbf{k}\in\text{BZ}}=\sum_{\mathbf{k}'}+\!\!\sum_{\mathbf{k'+M}}+\!\!\sum_{\mathbf{k'+K}}
+\!\!\sum_{\mathbf{k'+K+M}}+\!\!\sum_{\mathbf{k'-K}}+\!\!\sum_{\mathbf{k'-K+M}},
\end{equation}
where ${\mathbf{k}'\in\text{MBZ}}$ is implicitly assumed on the right hand side. 
Using the fact that ${3\mathbf{K}=\mathbf{b}_1-\mathbf{b}_2}$ and ${2\mathbf{M}=\mathbf{b}_2}$, 
we can express the whole mean-field Hamiltonian in MBZ as 
\begin{equation}
    H_\text{MF}=\sum_{\mathbf{k}\in\text{MBZ}}X_\mathbf{k}^\dagger H_\mathbf{k}X_\mathbf{k} 
    \label{MF1_k}
\end{equation}
with $X_\mathbf{k}=${\footnotesize $\left(f_{\mathbf{k}\uparrow},\,f_{\mathbf{k}\downarrow},\,f_{\mathbf{k+M}\uparrow},\,f_{\mathbf{k+M}\downarrow},\,f_{\mathbf{k+K}\uparrow},\,f_{\mathbf{k+K}\downarrow},\,f_{\mathbf{k+K+M}\uparrow},\right.$ $\left.\,f_{\mathbf{k+K+M}\downarrow},\,f_{\mathbf{k-K}\uparrow},\,f_{\mathbf{k-K}\downarrow},\,f_{\mathbf{k-K+M}\uparrow},\,f_{\mathbf{k-K+M}\downarrow}\right)^T$}, and
\begin{align}
    H_{\mathbf{k}}=H_{t,\mathbf{k}}+H_{SS,\mathbf{k}}&=\begin{bmatrix}
    H_{0}(\mathbf{k}) &  & \\
         & H_{0}(\mathbf{k+K}) & \\
         &  & H_{0}(\mathbf{k-K})
    \end{bmatrix}\nonumber\\
    &+\begin{bmatrix}
    V & H_1 & H_1^\dagger\\
        H_1^\dagger & V & H_1\\
        H_1 & H_1^\dagger & V
    \end{bmatrix}, \label{MF1_matrix}
\end{align}
where the blank entries refer to 0, and
\begin{align}
    H_{0}(\mathbf{k})&=2t\begin{bmatrix}
    A_\mathbf{k}  & B_\mathbf{k} 
    \vspace{2mm} \\
        B^*_\mathbf{k} & -A_\mathbf{k} 
\end{bmatrix}
 \otimes
{\mathcal I}_2
-\mu\,\mathcal{I}_{4},\\
V&=\tilde{J}_z \big[ \langle S_0^z\rangle+2\langle S_2^z\rangle \big]{\mathcal I}_2 \otimes \sigma_z 
,\\
H_1&=\tilde{J}_z\frac{\langle S_2^z\rangle-\langle S_0^z\rangle}{2}
{\mathcal I}_2 \otimes \sigma_z 
 -\tilde{J}_\perp \frac{\sqrt{3}\langle S_2^x\rangle}{2i}    
 {\mathcal I}_2 \otimes \sigma_x .
\end{align}
Here ${\mathcal I}_2$ and ${\mathcal I}_4$ are $2\times2$ and $4\times 4$ identity matrix, respectively. 
 By diagonalizing Eq.~(\ref{MF1_k}), we obtain 12 spinon bands in the MBZ as shown in the main text.

\section{Vanishing Chern number of filled spinon bands}\label{B}

With the quantum supersolid order, the Dirac cones at the Fermi level acquire the mass gap, 
and the 6 filled spinon bands under the Fermi level are well separated from the upper 6 spinon 
bands. As the time-reversal symmetry is already broken by the supersolid order, 
it is then natural to know whether the filled spinon bands support a nontrivial Chern number. 
When one goes beyond the mean-field theory, the gapped spinon bands with a nontrivial Chern number
could induce a Chern-Simon term in the U(1) gauge theory that is very much like the one 
in chiral spin liquids. 

Since all these bands are now overlapping in energy, one needs to invoke the non-Abelian 
Berry connection to define the Chern number of the filled bands. 
The Chern number of the total 6 filled spinon bands is well-defined 
by an integral over the continuum magnetic Brillouin zone (MBZ) as
\begin{equation}
C_{-}=\frac{1}{2\pi i}\int_\text{MBZ} \text{Tr} \, \big[\, d_{\mathbf{k}}\mathcal{A}^{-}_\mathbf{k} \big],
\end{equation}
where the non-Abelian Berry connection \cite{PhysRevLett.52.2111}
\begin{equation}
\mathcal{A}^{-}_\mathbf{k}={\psi^{-}_\mathbf{k}}^\dagger  {d}_\mathbf{k}\psi^{-}_\mathbf{k}
\end{equation}
 is a $6\times 6$ matrix-valued one-form in the momentum $\mathbf{k}$-space associated with the wavefunctions for the filled spinon bands $\psi^{-}_\mathbf{k}=(|\psi^{-}_{1\mathbf{k}}\rangle, ..., |\psi^{-}_{6\mathbf{k}}\rangle)$.

Numerically we evaluate this Chern number using the Fukui method \cite{doi:10.1143/JPSJ.74.1674}  
in the $M\times M$-grid discrete momentum space $k_l=(\frac{2\pi}{3Ma}j_x,\frac{\pi}{Ma\sqrt{3}}j_y)$ ($j_\mu=0,...,.M-1$ for $\mu=x,y$) as
\begin{equation}
    \tilde{C}_-=\frac{1}{2\pi i}\sum_l\tilde{F}_{xy}(k_l),
\end{equation}
where the lattice field strength 
\begin{equation}
\tilde{F}_{xy}(k_l)=\ln[U_x(k_l)U_y(k_l+\hat{x})U_x(k_l+\hat{y})^{-1}U_y(k_l)^{-1}] 
\end{equation}
with $\hat{x}=(\frac{2\pi}{3Ma},0)$ and $\hat{y}=(0,\frac{\pi}{Ma\sqrt{3}})$, and the link variable 
\begin{equation}
U_\mu(k_l)=\left|\text{det}\left[{\psi^-}^\dagger_{k_l}{\psi^-}_{k_l+\hat{\mu}}\right]\right|^{-1}\text{det}\left[{\psi^-}^\dagger_{k_l} {\psi^-}_{k_l+\hat{\mu}}\right].
\end{equation}
 With the periodic boundary condition, 
 the MBZ is a closed two-dimensional torus $T^2$, guaranteeing the Chern number an integer. 
 Then, with sufficiently fine grids, the Chern number on the discretized space $\tilde{C}_-$ equals to $C_-$, 
 and we find ${\tilde{C}_-=0}$ up to a grid choice ${M=300}$. 
 Thus, we conclude that the gap opening from the precursory DSL 
 with the supersolid order is not topological and there is no Chern-Simons term.

\section{Effect of external magnetic field}\label{C}
\begin{figure*}[ht]
\centering
\includegraphics[width=1\textwidth]{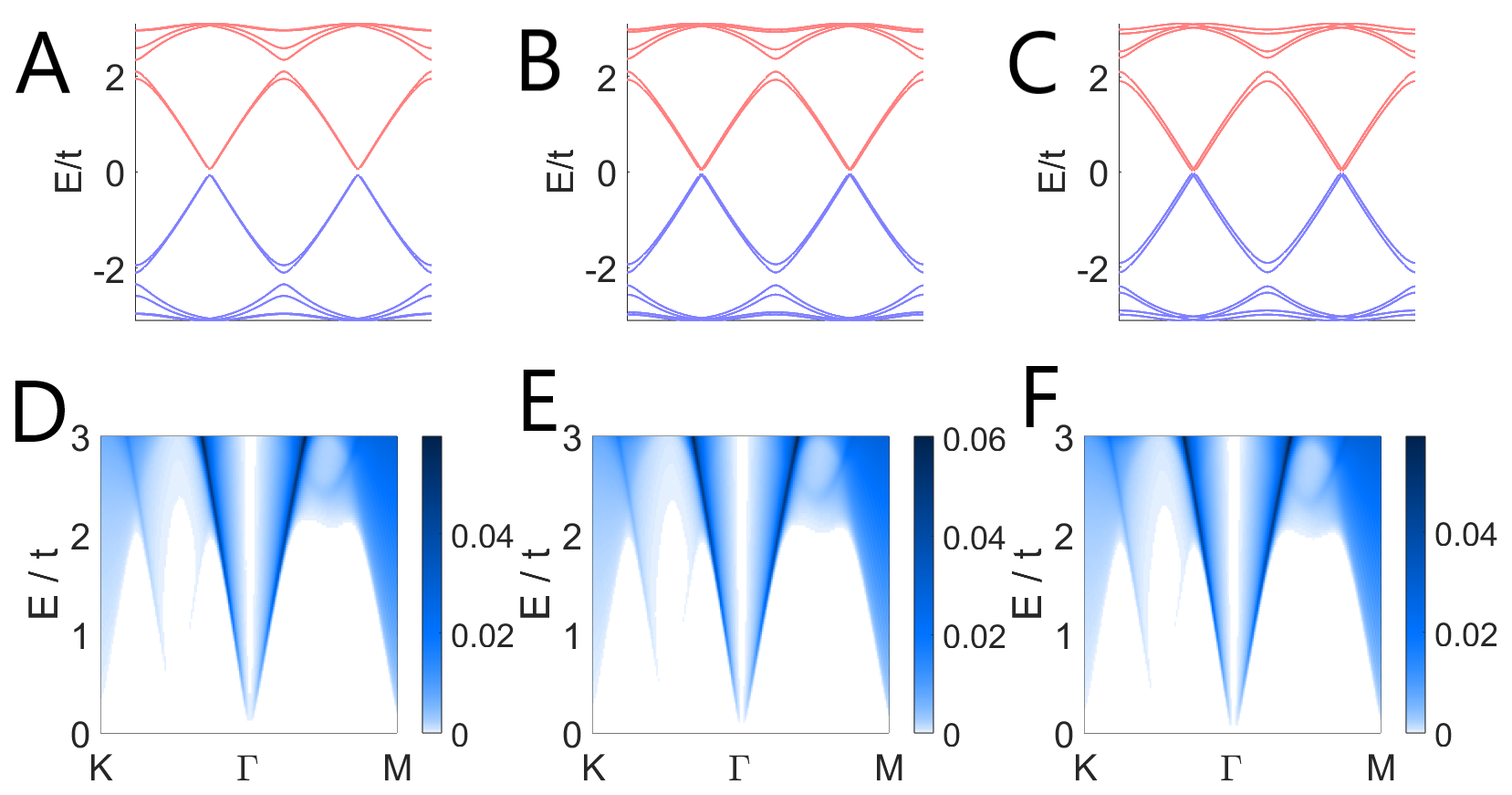}
\caption{The calculated spinon spectrum and INS signal under various weak perpendicular magnetic field. 
The external field polarizes the magnetic moments and induces a net magnetization along $z$ axis. 
We take the magnetic susceptibility ${{\rm d}M/{\rm d}B = 1\mu_B}$T$^{-1}/{\rm Co}$~\cite{sheng2022two}. 
The subfigures (a) and (d)
show the result with ${{\rm B}_z = 0}$T 
(${\theta = 57.4^{\circ}}$), while (b,e) and (c,f) are those with ${{\rm B}_z = 0.1t}$ (${\approx 0.0117 {\rm T}}$, ${\theta = 63.7^{\circ}}$),
${{\rm B}_z = 0.2t}$ (${\approx 0.0234 {\rm T}}$, ${\theta = 72.0^{\circ}}$), respectively.  
Notice that the predicted continuum signal at the $\mathbf{K}$ and $\mathbf{M}$ points persists.} 
\label{fig: fig9}
\end{figure*}

Here we further examine our model with the presence of a weak external magnetic field.
For the in-plane magnetic field, the continuous U(1) symmetry will be absent, and the Goldstone mode will
disappear. The magnetic excitation will be fully gapped. For out-of-plane field with ${{\mathbf B} = B\hat{z}}$ 
along the $z$ direction, the continuous U(1) symmetry is still present in the model. 
Although the field slightly polarizes the spins,
as long as the field is perturbative on the supersolid order, the Goldstone mode 
of the spin-wave excitation should persist in the weak field regime. 
For the spinon continuum, we do not expect a significant change as well, 
and the continuous excitations should persist in the weak field regime. 
In Fig.~\ref{fig: fig9}, we computed the INS results 
for the spinon continuum under the weak field. 
We have let the magnetic moments on the red and green sublattices 
to rotate while keeping their magnitude fixed. 
The net magnetic susceptibility is ${\rm d}M/{\rm d}B = 1\mu_B$T$^{-1}/{\rm Co}$~\cite{sheng2022two}. 
We find that, the addition of ${\mathbf B}$ only slightly affects the spinon spectrum. 
One can understand this by reviewing the statements that were made in Fig.~\ref{fig: fig4}, 
where the addition of $\tilde{J}_z$ splits up the spin-up and spin-down bands. 
Here the external field in the $z$ direction does a similar thing, that is to further split them, 
while the gap size is not much affected since it is determined by $\tilde{J}_{\perp}$. 
As for the INS spectrum, the effect of external $B_z$ is almost negligible in our calculation. 
The predicted spectral continuum at the K and M points still exists. 

\section{Finite temperature behaviors}\label{D}

\begin{figure}[b]
\centering
\includegraphics[width=7.5cm]{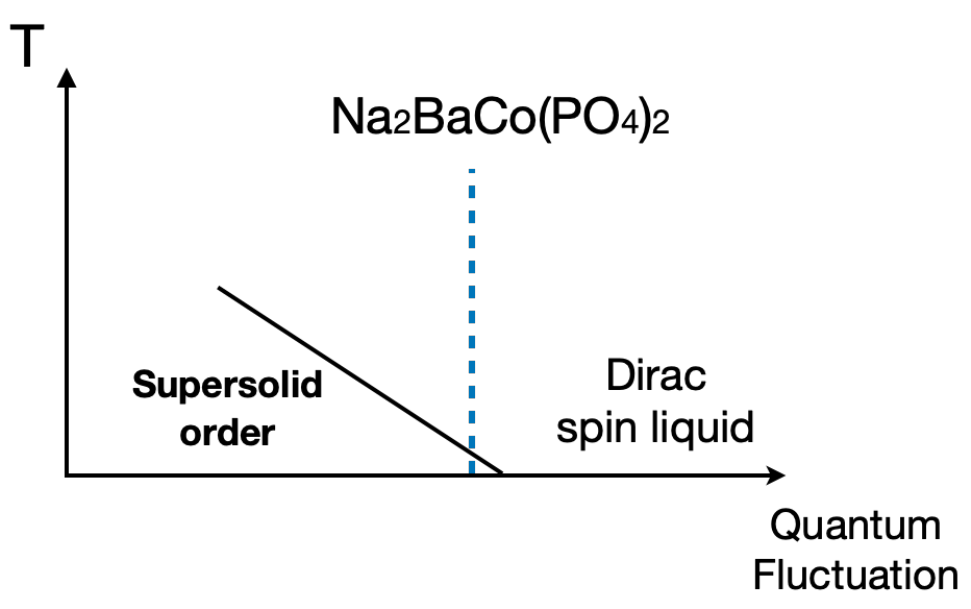}
\caption{The expected finite temperature behaviors, and the dashed line is for Na$_2$BaCo(PO$_4$)$_2$. 
The system has a weak supersolid order at very low temperature. 
Thermal fluctuation suppresses the order and drives the system into the Dirac spin liquid regime.} 
\label{fig: fig10}
\end{figure}

Since the spinon gap is induced by the supersolid magnetic order, when the thermal fluctuations 
suppress the magnetic order at low and finite temperature, the system would cross over to the Dirac spin liquid regime. 
Although this is the finite temperature paramagnetic phase, the physical properties are governed by the Dirac spin liquid.
The expected finite-temperature phase diagram is depicted in Fig.~\ref{fig: fig10}, where the increased 
quantum fluctuation would further stabilize the spin liquid state at zero temperature.  The similar picture
was actually invoked for $\alpha$-RuCl$_3$~\cite{Banerjee_2016} that also orders at low temperature. 
Thus, at the temperature above the ordering, the (gapless) Dirac spinons should give a heat capacity of $T^2$ and 
contribute to the thermal transport. As we have mentioned that, Ref.~\onlinecite{NBCP_magnetization} 
already found the signature of itinerant excitations. 
Moreover, Ref.~\onlinecite{thermalHall} further proprosed the gapless spin liquid from the finite temperature
thermal Hall transport result. Although the thermal Hall result is not inconsistent with a gapless spin liquid,
the generation of Berry curvature distribution for the spinons requires further understanding.

\section{More discussion about quantum spin supersolid and Dirac fermions}

Since Na$_2$BaCo(PO$_4$)$_2$ realizes the quantum spin supersolid 
at low temperatures and we have further proposed a precursory Dirac spin liquid for it,
it is worthwhile to raise some further discussion about the proximate states out of 
the quantum spin supersolid and some of the related properties. 
This may provide some further insights for the interested readers. 
The material Na$_2$BaCo(PO$_4$)$_2$ is located in the fully frustrated 
regime of the XXZ model on the triangular lattice and thus has a sign 
problem for quantum Monte Carlo simulation. 
Most of the early analysis of the triangular lattice XXZ model 
was devoted to the less frustrated regime of the XXZ model, {i.e. } ${J_{\perp} <0}$
where the quantum supersolid is well present before the knowledge 
on the fully frustrated side due to the absence of the sign problem~\cite{melko2005supersolid,wessel2005supersolid}. 
The intimate relation between the fully frustrated regime with ${J_{\perp} >0}$ and 
the less frustrated regime with ${J_{\perp} <0}$ was partially addressed in
Ref.~\onlinecite{PhysRevLett.102.017203}.
It was shown that~\cite{PhysRevLett.102.017203}, in the perturbative regime ${J_{\perp} \ll J_z}$,
one can perform a unitary transformation to connect both regimes. 
This establishes the presence of the supersolid in the fully frustrated regime
and argued to extend to the Heisenberg point based on the continuity argument. 
This unitary transformation that connects the fully frustrated and less frustrated regime 
is analogous to the one that was used in the pyrochlore spin ice context~\cite{PhysRevB.69.064404}. 

 On the less frustrated side with ${J_{\perp} <0}$, 
 due to the underlying U(1) symmetry of the XXZ model, 
 a boson-vortex duality was adopted to explore the proximate 
 states and phase transitions by viewing the spins as hardcore bosons~\cite{PhysRevB.72.134502}. 
 It was predicted from an emergent SU(2) symmetry that,
 the quantum supersolid is proximate to deconfined 
 quantum criticality of the NCCP$^1$ universality class. 
 Remarkably, this is identical to the N\'{e}el-VBS transition. 
 From the recent duality argument, this criticality 
 is dual to the  
${N_f=2}$ fermionic quantum electrodynamics
with Dirac fermions~\cite{PhysRevX.7.031051}. 
Although this is on the less frustrated side and may not be
directly related to Na$_2$BaCo(PO$_4$)$_2$, 
it provides some insights 
about the connection between the quantum supersolid and 
the Dirac fermions, and may be useful
for the full understanding of the nearby phases of the 
quantum supersolids in both regimes.

\bibliography{NBCP}

\end{document}